\documentclass[nohyper]{JHEP3}

\usepackage{amsmath,amssymb}
\usepackage[dvips]{graphicx}
\usepackage{afterpage}
\usepackage{bbm}

\newcommand{\ie}{\textit{i.e.}}
\newcommand{\eg}{\textit{e.g.}}
\newcommand{\LT}{\left}
\newcommand{\RT}{\right}
\newcommand{\eVq}{\ensuremath{\text{eV}^2}}
\newcommand{\Dmq}{\Delta m^2}
\newcommand{\diag}{\mathop{\mathrm{diag}}}
\renewcommand{\Re}{\mathop{\mathrm{Re}}}
\renewcommand{\Im}{\mathop{\mathrm{Im}}}

\newenvironment{myitemize}
{\begin{list}{\textbullet}{\setlength{\leftmargin}{\parindent}}}
{\end{list}}

\newcommand{\textitem}[1]{\par\bigskip\noindent\textbf{#1}}

\newcommand{\PAGEFIGURE}[1]{\FIGURE[!p]{#1}\afterpage\clearpage}

\title{1-3 leptonic mixing and the neutrino oscillograms of the
  Earth}

\author{Evgeny Kh.~Akhmedov%
  \footnote{on leave from the National Research Centre Kurchatov
    Institute, Moscow, Russia}\\
  Department of Theoretical Physics,
  Royal Institute of Technology (KTH),\\
  AlbaNova University Center, SE-106 91 Stockholm, Sweden\\
  E-mail: \email{akhmedov@mpi-hd.mpg.de}}

\author{Michele Maltoni\\
  The Abdus Salam International Centre for Theoretical Physics,\\
  Strada Costiera 11, I-34014 Trieste, Italy\\
  {\rm and:}
  Departamento de F\'isica Te\'orica \& Instituto de F\'isica Te\'orica,\\
  Facultad de Ciencias C-XI, Universidad Aut\'onoma de Madrid,\\
  Cantoblanco, E-28049 Madrid, Spain\\
  E-mail: \email{maltoni@delta.ft.uam.es}}

\author{Alexei Yu.~Smirnov\\
  The Abdus Salam International Centre for Theoretical Physics,\\
  Strada Costiera 11, I-34014 Trieste, Italy\\
  {\rm and:}
  Institute for Nuclear Research, Russian Academy of Sciences\\
  Moscow, Russia\\
  E-mail: \email{smirnov@ictp.trieste.it}}

\abstract{%
  We develop a detailed and comprehensive description of neutrino
  oscillations driven by the 1-3 mixing in the matter of the Earth.
  The description is valid for the realistic (PREM) Earth density
  profile in the whole range of nadir angles and for neutrino energies
  above 1 GeV. It can be applied to oscillations of atmospheric and
  accelerator neutrinos. The results are presented in the form of
  neutrino oscillograms of the Earth, \ie\ the contours of equal
  oscillation probabilities in the neutrino energy--nadir angle plane.
  A detailed physics interpretation of the oscillograms, which
  includes the MSW peaks, parametric ridges, local maxima, zeros and
  saddle points, is given in terms of the amplitude and phase
  conditions. Precise analytic formulas for the probabilities are
  obtained.
  We study the dependence of the oscillation pattern on $\theta_{13}$
  and find, in particular, that the transition probability $P > 1/2$
  appears for $\sin^2 2\theta_{13}$ as small as $\sim 0.009$.  We
  consider the dependence of the oscillation pattern on the matter
  density profile and comment on the possibility of the oscillation
  tomography of the Earth.}

\preprint{IFT-UAM/CSIC-07-06\\
  hep-ph/0612285}


\begin{document}


\section{Introduction}

Substantial future progress in neutrino physics will be related to the
long baseline experiments as well as studies of the cosmic and
atmospheric neutrinos. The key element of these experiments is that
neutrinos propagate long distances inside the Earth before reaching
detectors. Oscillations in the matter of the Earth change flavor
properties of neutrino fluxes, which opens up possibilities to 
\begin{myitemize}
  \item study dynamics of various oscillation effects; 
    
  \item determine yet unknown oscillation parameters: 1-3 mixing,
    deviation of the 2-3 mixing from maximal and its octant, the type
    of the neutrino mass hierarchy, and CP-violation;
    
  \item perform the oscillation tomography of the Earth;
    
  \item search for various new effects caused by non-standard neutrino
    interactions and/or by exotic physics (such as violation of
    Lorentz or CPT invariance).
\end{myitemize}

The sensitivity of present-day experiments to the corresponding 
effects, that appear usually as small corrections to the leading
oscillation phenomena (\eg\ $\nu_\mu \leftrightarrow \nu_\tau$ vacuum
oscillations), is rather low. In the case of atmospheric neutrinos
this is related to low fluxes and therefore low statistics of events,
and also to uncertainties in the fluxes.  The accelerator neutrino
beams are, in principle, well controlled, however in majority of the
projects the baselines are relatively short, whereas the very long
baseline experiments (\eg\ with neutrino factories) employ high energy
neutrinos. This means that the region of energy $E_\nu \sim (3 - 10)$
GeV and nadir angle $\Theta_\nu = 0 - 70^\circ$, where the most
interesting oscillation phenomena occur, is not covered. Therefore one
needs to rely on high precision measurements of the suppressed
``tails'' of these phenomena.  The dilemma is whether to study small
effects with a number of systematical errors and degeneracies, or
develop experimental approaches that will allow to cover the regions
of large oscillation effects.  The latter option is realized in the 
concepts of very long baseline accelerator experiments and very large
volume atmospheric neutrino detectors. 

In connection with possible future experimental developments and
discussions of new strategies of research, a detailed and
comprehensive study of physics of the neutrino oscillations in the
Earth is needed. Some studies in this direction have been carried out
in the past. 

In the matter of the Earth, the resonance enhancement of $\nu_e
\leftrightarrow \nu_\mu$ and $\nu_\tau$ oscillations can take
place~\cite{Wolfenstein:1977ue, Mikheev:1986gs}, with the MSW
resonance peak at $E_\nu / \Dmq \sim 2.5 \times 10^{3}~\text{GeV} /
\eVq$~\cite{Smirnov:1986ij, resenh}.  It was also recognized that the
size of the Earth is comparable with the neutrino refraction length,
and consequently strong enhancement of oscillations can occur for
rather deep trajectories (small nadir angles) and not too small mixing
angles: $\sin^2 2 \theta_{13} > 0.03$.  With decreasing baseline,
first the matter effect and then the oscillation effect disappear.
This phenomenon ~\cite{Wolfenstein:1977ue,vacmim}, termed ``vacuum
mimicking'', implies that for short baselines the flavour transitions
are described by the vacuum oscillation formula.

For 1-2 mixing the resonance is in the neutrino channel and at rather
low energies: $E_\nu \sim 0.1$ GeV.
For non-zero 1-3 mixing the MSW resonance peak related to the
atmospheric mass splitting is at $E_\nu \sim 6$ GeV.  The peak is
narrow and, depending on the mass hierarchy, the enhancement occurs in
the neutrino (normal hierarchy) or antineutrino (inverted hierarchy) 
channels.  Possibilities to observe this resonance peak have been
explored in connection to the measurements of the mixing angle
$\theta_{13}$ and determination of the neutrino mass
hierarchy~\cite{13res, 13res-vla, 13res-vlb, 13res-vlc, 13res-h0,
13res-h1, 13res-h2}. 
Furthermore, it was realized that matter effects can also strongly
influence $\nu_\mu \leftrightarrow \nu_\tau$
oscillations~\cite{Freund:1999vc, mutau}.

A qualitatively new oscillation phenomenon can be realized for 
neutrinos crossing the core of the Earth due to a sharp change of the
density at the border between the core and the mantle~\cite{LS, P,
Akhm2, ADLS, CMP, CP}. In particular, for non-zero 1-3 mixing, the
existence of an additional peak is predicted in the energy
distribution between the peaks due to the MSW resonances in the core
and mantle.
This core-mantle effect was interpreted as being due to the parametric
enhancement of neutrino oscillations~\cite{LS, Akhm2,
ADLS}.\footnote{The parametric resonance in neutrino oscillations in
the Earth was first discussed for $\nu_\mu \leftrightarrow \nu_s$
oscillations of atmospheric neutrinos in~\cite{LS}, though the 
parametric peak appeared in some early numerical 
computations~\cite{Smirnov:1986ij, resenh}.} The additional peak is a
manifestation of the parametric resonance for 3 layers (1.5 period of
the ``castle wall'' profile) in matter~\cite{Akhmedov:1999va}.  The
parametric resonance occurs when the variation of the matter density
along the neutrino trajectory is in a certain way correlated with the
values of the oscillation parameters~\cite{ETC, Akhm1, KS}.
A different interpretation of this peak, as being due to an
interference between the contributions to the oscillation amplitude
from different layers of the Earth's mantle and core, has been
discussed in~\cite{P, CMP, CP}.  It was uncovered~\cite{AMS1} that the
parametric resonance condition is fulfilled also at large energies
(above the mantle MSW resonance), leading to the appearance of two
parametric peaks in the nadir angle distribution.  For general
consideration of evolution in the multi-layer media, 
see~\cite{layers}.

Apart from the large scale structures of the density profile, effects
of small scale density perturbation have been explored~\cite{dpertur}.

Analytic approaches have been developed to describe the physics of
various oscillation effects, to understand the influence of the
density profile modifications on the oscillation probabilities and
also to simplify numerical computations.  Many studies have been
performed in the constant density approximation~\cite{Smirnov:1986ij,
resenh} or approximation of several layers of constant
densities~\cite{LS, P, Akhm2, ADLS, CMP, CP, Akhm1, KS, 3nuconst,
Freund:1999vc,cm-applic}. For the varying densities within the layers
a perturbation-theory approach has been developed in
\cite{Lisi:1997yc} for the description of the solar neutrino
oscillations in the Earth. 

For $3\nu$ mixing the analytic approaches employ various expansions in
the small parameters $\sin^2 \theta_{13}$ and/or in the ratio of the
mass squared differences $\Dmq_{21} / \Dmq_{31}$~\cite{expansion}. For
the case of non-constant density matter, different perturbation theory
approaches have been developed for oscillations in low
density~\cite{RoyC, approx-low} and high density~\cite{approx-high,
AMS1} media. In~\cite{AMS1} an analytic description of oscillations in
the high energy limit has been worked out for the case of the
realistic (PREM) density profile.  
The influence of the Earth density profile on the oscillation
probabilities has also been considered in~\cite{tomog}. 
 
To some extent, the results obtained so far had fragmentary character.
This paper is the first one in a series of papers we devote to a
detailed and comprehensive study of oscillations of neutrinos inside
the Earth. Here we present a thorough description of neutrino
oscillations caused by non-zero 1-3 mixing, with the emphasis on the
physics of the phenomenon. In particular, we study the complex pattern
and interplay of various oscillation resonances.  We show that the
main features of this complex pattern can be easily understood in
terms of different realizations of just two simple conditions -- the
amplitude condition and the phase condition. We perform our study in
terms of ``neutrino oscillograms'' of the Earth: contours of constant
oscillation probabilities in the plane of neutrino energy and nadir
angle. These plots have been presented in several earlier publications
as illustrations~\cite{CMP, 13res-vlb, oscillograms}. Here we use them
as the main tool of the study.  We present a description of all the
structures of the oscillograms and of their dependence on
$\theta_{13}$ and the density profile.  We generalize the amplitude
and the phase conditions to the case of non-constant densities in the
layers of matter and develop an appropriate analytic formalism.  

The paper is organized as follows. In Sec.~\ref{sec:neutrosc}, after
some generalities, we introduce the neutrino oscillograms of the
Earth, concentrating mainly on $\nu_e \leftrightarrow
\nu_{\mu}(\nu_{\tau})$ transitions. We discuss the accuracy of the
constant-density layers approximation of the Earth matter profile and
describe the graphical representation of the conversion. In
Sec.~\ref{sec:interposc} we give the physics interpretation of the
oscillograms. In Sec.~\ref{sec:approx} we derive approximate analytic
formulas for probabilities for a realistic matter density profile. In
Sec.~\ref{sec:channels} we study the dependence of the oscillograms on
the 1-3 mixing and the density profile of the Earth, and we discuss
the oscillation probabilities for the other channels. Discussion and
conclusions follow in Sec.~\ref{sec:conclusions}.


\section{Neutrino oscillograms of the Earth}
\label{sec:neutrosc}


\subsection{Context. Evolution matrix}
\label{sec:3fosc}

We consider the three-flavor neutrino system with the state vector
$\nu_f \equiv (\nu_e,\, \nu_{\mu},\, \nu_{\tau})^T$. Its evolution in
matter is described by the equation
\begin{equation} \label{eq:evolution}
    i \frac{d \nu_f}{dx} =
    \LT( \frac{U M^2 U^\dagger}{2 E_\nu} + \hat{V}(x) \RT) \nu_f\,,
\end{equation}
where $E_\nu$ is the neutrino energy and $M^2 \equiv \diag(0,\,
\Dmq_{21},\, \Dmq_{31})$ is the diagonal matrix of neutrino mass
squared differences. $\hat{V}(x) \equiv \diag(V(x),\, 0,\, 0)$ is the
matrix of matter-induced neutrino potentials with $V(x) \equiv \sqrt 2
G_F N_e(x)$, $G_F$ and $N_e(x)$ being the Fermi constant and the
electron number density, respectively. The mixing matrix $U$ defined
through $\nu_f = U \nu_{m}$, where $\nu_{m} = (\nu_1,\, \nu_2,\,
\nu_3)^T$ is the vector of neutrino mass eigenstates, can be
parameterized as
\begin{equation} \label{eq:mixing}
    U = U_{23} I_{\delta} U_{13} I_{-\delta}U_{12} \,.
\end{equation}
Here the matrices $U_{ij} = U_{ij}(\theta_{ij})$ describe rotations in
the $ij$-planes by the angles $\theta_{ij}$, and $I_{\delta} \equiv
\diag(1,\, 1,\, e^{i\delta})$, where $\delta$ is the Dirac-type
CP-violating phase.

Let us introduce the evolution matrix (the matrix of transition and
survival amplitudes) $S(x, x_0)$, which describes the evolution of the
neutrino state over a finite distance: from $x_0$ to $x$. To simplify
the presentation, throughout the paper we will use the notation $S(x)
\equiv S(x, 0)$ and $S \equiv S(L)$, where $L$ is the total length of
the trajectory. The matrix $S(x)$ satisfies the same evolution
equation as the state vector~\eqref{eq:evolution}:
\begin{equation} \label{eq:smatreq}
    i \frac{dS(x)}{dx} = H(x) \, S(x) \,.
\end{equation}
The solution of equation~\eqref{eq:smatreq} with the initial condition
$S(0)=\mathbbm{1}$ can be formally written as
\begin{equation}
    S(x) = T \exp \LT(- i \int_0^x  H \, dx \RT)\,.
\end{equation}

It is convenient to consider the evolution of the neutrino system in
the propagation basis, $\tilde{\nu} = (\nu_e,\, \tilde{\nu}_2,\,
\tilde{\nu}_3)^T$, defined through the relation
\begin{equation} \label{eq:basisrel}
    \nu_f = U_{23}I_{\delta} \tilde{\nu} \,.
\end{equation}
As follows from~\eqref{eq:evolution} and~\eqref{eq:mixing}, the
Hamiltonian $\tilde{H}$, that describes the evolution of the neutrino
vector of state $\tilde{\nu}$, is
\begin{equation}
    \tilde{H}(x) = \frac{1}{2 E_\nu} U_{13} U_{12} M^2
    U^\dagger_{12} U^\dagger_{13} + \hat{V}(x) \,.
\end{equation}
This Hamiltonian does not depend on the 2-3 mixing and CP-violating
phase. The dependence on these parameters appears when one projects
the initial state on the propagation basis and the final state back
onto the original flavor basis. Explicitly, the Hamiltonian
$\tilde{H}$ reads
\begin{equation} \label{eq:matr1}
    \tilde{H}(x) =
    \frac{\Dmq_{31}}{2 E_\nu}
    \begin{pmatrix}
	s_{13}^2 + s_{12}^2\, c_{13}^2\,r_\Delta + 2V(x) E_\nu / \Dmq_{31}
	& s_{12}\,c_{12}\,c_{13}\,r_\Delta
	& s_{13}\,c_{13}(1 - s^2_{12}\,r_\Delta)
	\\
	\ldots
	& c_{12}^2\,r_\Delta
	& - s_{12}\,c_{12}\,s_{13}\,r_\Delta
	\\
	\ldots
	& \ldots
	& c_{13}^2 + s_{12}^2\,s_{13}^2\,r_\Delta
    \end{pmatrix}
\end{equation}
where $c_{ij} \equiv \cos \theta_{ij}$, $s_{ij} \equiv \sin
\theta_{ij}$ and
\begin{equation}
    r_\Delta \equiv \frac{\Dmq_{21}}{\Dmq_{31}} \,.
\end{equation}
According to~\eqref{eq:basisrel}, the evolution matrix $\tilde{S}(x)$
in the basis $(\nu_e, \tilde{\nu}_2, \tilde{\nu}_3)$ is related to
$S(x)$ by the transformation:
\begin{equation}
    S(x) = \tilde{U} \, \tilde{S}(x) \, \tilde{U}^{\dagger},
    \qquad \tilde{U} \equiv U_{23}I_{\delta}.
\end{equation}
The evolution of $\tilde{S}(x)$ is given by the equation analogous to 
Eq.~\eqref{eq:smatreq} with the Hamiltonian $\tilde{H}(x)$.


\subsection{High energy neutrino approximation}
\label{sec:heosc}

For sufficiently high energies ($E_\nu > 1-2$ GeV) one can neglect the
1-2 mass splitting. Then, according to~\eqref{eq:matr1}, the state
$\tilde{\nu}_2$ decouples from the rest of the system and does not
evolve. Therefore, if we parameterize $\tilde{S}$ in the basis
$(\nu_e, \tilde{\nu}_2, \tilde{\nu}_3)$ as
\begin{equation} \label{eq:matr2}
    \tilde{S} =
    \begin{pmatrix}
	A_{ee} & A_{e2} & A_{e3} \\
	A_{2e} & A_{22} & A_{23} \\
	A_{3e} & A_{32} & A_{33}
    \end{pmatrix}
\end{equation}
we find that in this approximation $A_{e2} = A_{2e} = A_{23} = A_{32}
= 0$, $A_{22} = 1$ and the evolution matrix in the flavor basis takes
the form
\begin{equation} \label{eq:matr3}
    S \approx
    \begin{pmatrix}
	A_{ee} & s_{23} A_{e3}  & c_{23}A_{e3}
	\\
	s_{23} A_{3e} & c_{23}^2 A_{22} + s_{23}^2 A_{33} &
	- s_{23}c_{23} (A_{22} - A_{33})
	\\
	c_{23} A_{3e}  &  - s_{23}c_{23} (A_{22} - A_{33}) &
	s_{23}^2 A_{22} + c_{23}^2 A_{33}
    \end{pmatrix} \,,
\end{equation}
where we omitted the CP-phase factor $e^{-i\delta}$ since CP-violating
effects are absent in the limit $r_\Delta \to 0$. We will consider the
complete $3\nu$ system in the next publication~\cite{AMS2}.
Denoting
\begin{equation}
    P_A \equiv |A_{e3}|^2 = |A_{3e}|^2 \,,
\end{equation}
we obtain from~\eqref{eq:matr3}
\begin{align}
    \label{eq:prob-ee}
    P(\nu_e\to \nu_e) &= 1 - P_A,
    \\
    \label{eq:prob-emu}
    P(\nu_e\to \nu_\mu) &= P(\nu_\mu\to \nu_e) = s_{23}^2  P_A \,,
    \\
    \label{eq:prob-etau}
    P(\nu_e\to \nu_\tau) &=  P(\nu_\tau \to \nu_e) = c_{23}^2 P_A \,,
    \\
    \label{eq:prob-mumu}
    P(\nu_\mu\to\nu_\mu ) &= 1  - s_{23}^4 P_A  -
    2s_{23}^2 c_{23}^2\,\LT[1 - \Re A_{33} \RT] \,,
    \\
    \label{eq:prob-mutau}
    P(\nu_\mu\to\nu_\tau ) &=   -  s_{23}^2\,c_{23}^2\, P_A
    + 2s_{23}^2 c_{23}^2\,\LT[1 - \Re A_{33} \RT] \,.
\end{align}
The formulas in Eqs.~(\ref{eq:prob-ee}--\ref{eq:prob-mutau}) reproduce
the probabilities derived in~\cite{ADLS}.

Thus, in the approximation $\Dmq_{21} = 0$ the dynamical problem is
reduced to two flavor evolution problem.  Throughout this paper we
work in the basis where, for a 2-flavor neutrino system, $H_{11} =
-H_{22} = -\cos 2 \theta_{13}\Dmq/4E + V/2$.  This can be always
achieved by subtracting from $H$ a matrix proportional to the unit 
matrix and correspondingly rephasing the neutrino vector of state.
This symmetric Hamiltonian is related to the Hamiltonian of the 1-3
subsystem $\tilde{H}^{(13)}$, obtained from Eq.~\eqref{eq:matr1} by
taking the limit $r_\Delta=0$ and removing the decoupled state
$\tilde{\nu}_2$, as
\begin{equation}
    \tilde{H}^{(13)} = H + \LT(\frac{\Dmq}{4E_\nu} +
    \frac{V}{2}\RT)\mathbbm{1} \,.
    \label{eq:HHconnect}
\end{equation}
For the $2\nu$ case the unitary evolution matrix can be parameterized
as
\begin{equation} \label{eq:U1}
    S =
    \begin{pmatrix}
	\hphantom{-}\alpha & \beta \\
	-\beta^* & \alpha^*
    \end{pmatrix}\,,
    \qquad |\alpha|^2+|\beta|^2 = 1.
\end{equation}
For density profiles that are symmetric with respect to the midpoint
of the neutrino trajectory (for brevity, symmetric profiles), T
invariance leads to the equality of the off-diagonal elements of the
evolution matrix, $\beta = -\beta^*$, which means that $\beta$ is pure
imaginary~\cite{Tviol}.

For a single layer of constant density the solution can be written
explicitly:
\begin{equation} \label{eq:Ubar}
    S(x) =
    \begin{pmatrix}
	\cos\phi(x) + i \cos2 \theta_m \, \sin\phi(x)
	& -i \sin2 \theta_m \, \sin\phi(x) \\
	-i \sin2 \theta_m \, \sin\phi(x)
	& \cos\phi(x) - i \cos2 \theta_m \, \sin\phi(x)
    \end{pmatrix},
    \quad
    \phi(x) \equiv \bar\omega \, x \,.
\end{equation}
Here is $\theta_m$ is the mixing angle in matter and $\phi(x)$ is the
half-phase of oscillations in matter with
\begin{equation}
    \bar\omega = \omega(\bar{V}) \equiv
    \sqrt{\LT( \cos 2\theta_{13} \frac{\Dmq_{31}}{4 E_\nu}
      - \frac{\bar{V}}{2} \RT)^2
      + \LT(\sin 2\theta_{13} \frac{\Dmq_{31}}{4 E_\nu} \RT)^2}.
\end{equation}
The moduli squared of the elements of ${S}$ reproduce the well-known
probabilities for oscillations in a uniform medium. Thus in the
notation of Eq.~\eqref{eq:U1},
\begin{equation} \label{eq:ab1}
    \alpha = \cos \phi + i \cos 2\theta_m \, \sin \phi  \,,\qquad
    \beta = -i \sin 2\theta_m \, \sin \phi \,.
\end{equation}
In what follows we will generalize this result to the cases of several
layers of constant densities and also of changing densities within the
layers. 

The scale of the matter effects in neutrino oscillations is set up by
the refraction length, $l_0 \equiv 2\pi/(\sqrt{2}G_F N_e)$, which is
comparable with the radius of the Earth $R = 6371$ km. The oscillation
length in matter $l_m$ is given by
\begin{equation}
    \frac{2\pi}{l_m} = 2\bar\omega \,.
\end{equation}
At high energies the length $l_m$ essentially coincides with the
refraction length: $l_m \simeq l_0$. In the resonance channel, with
decreasing energy $l_m$ increases and reaches its maximum slightly
above the resonance energy. At the MSW resonance, $E_\nu = E_R$, one 
has $l_m = l_\nu / \sin 2 \theta_{13}$, where $l_\nu$ is the vacuum 
oscillation length. Below the resonance, with energy further
decreasing, $l_m$ decreases and approaches $l_\nu$: $l_m(E_\nu \ll
E_R)\approx l_\nu$. 

In a non-uniform density medium the adiabatic half-phase of
oscillations is defined as 
\begin{equation} \label{eq:adphase}
    \phi(x) \equiv \int_0^x \omega(x') \, dx' \,, \qquad
    \omega(x) \equiv \omega(V(x)) \,.
\end{equation}


\subsection{Neutrino oscillograms of the Earth}
\label{sec:oscillogr}

For given values of $|\Dmq_{31}|$, $\sin^2 2\theta_{13}$, and the type
of neutrino mass hierarchy, the oscillation probabilities depend on
the neutrino energy $E_\nu$ and the nadir angle of its trajectory
$\Theta_\nu$. Therefore a complete description of the oscillation
pattern can be given by contours of equal oscillation probabilities in
the $(E_\nu ,\, \Theta_\nu)$ plane. We call the resulting figures the
\emph{neutrino oscillograms} of the Earth. The plots of this type were
produced for the first time by P.~Lipari in 1998 (unpublished) and
then appeared in several later publications~\cite{CMP, 13res-vlb,
oscillograms}. Here we will employ this kind of plots as the main tool
of our study.

We use in our calculations the matter density distribution inside the
Earth, $\rho$, as given by the PREM model~\cite{PREM}. It exhibits a
characteristic structure with a relatively slow density change within
the mantle and within the core and a sudden jump of density by about a
factor of two at their border. Smaller jumps appear between the inner
core and the outer one and near the surface of the Earth. The number
of electrons per nucleon, $Y_e \equiv N_e m_N/\rho$ where $m_N$ is the
nucleon mass, equals $Y_e = 0.497$ in the mantle and $Y_e = 0.468$ in
the core. For the energies $E_\nu > (1-2)$ GeV the oscillation length
in vacuum $l_\nu$ exceeds 1000 km, and therefore the effects of
smaller-scale density perturbations are averaged out.
Furthermore, the density profile experienced by neutrinos along any
trajectory inside the Earth is symmetric with respect to the midpoint
of the trajectory,\footnote{We neglect possible short-scale
inhomogeneities of the matter density distribution in the Earth which
may violate this symmetry.} so that
\begin{equation}
    V(x) = V(L - x),
\end{equation}
where $L = 2R \cos\Theta_\nu$ is the length of the trajectory. As we
will see below, to a large extent this feature determines the
properties of the oscillation probabilities and the structure of the
oscillograms.

For numerical calculations we use the current best-fit value of
$\Dmq_{31} = 2.5 \times 10^{-3}$ eV; in the case of negligible effects
of 1-2 splitting, changing $\Dmq_{31}$ is equivalent to
correspondingly rescaling the neutrino energy.

\PAGEFIGURE{
  \includegraphics[width=145mm]{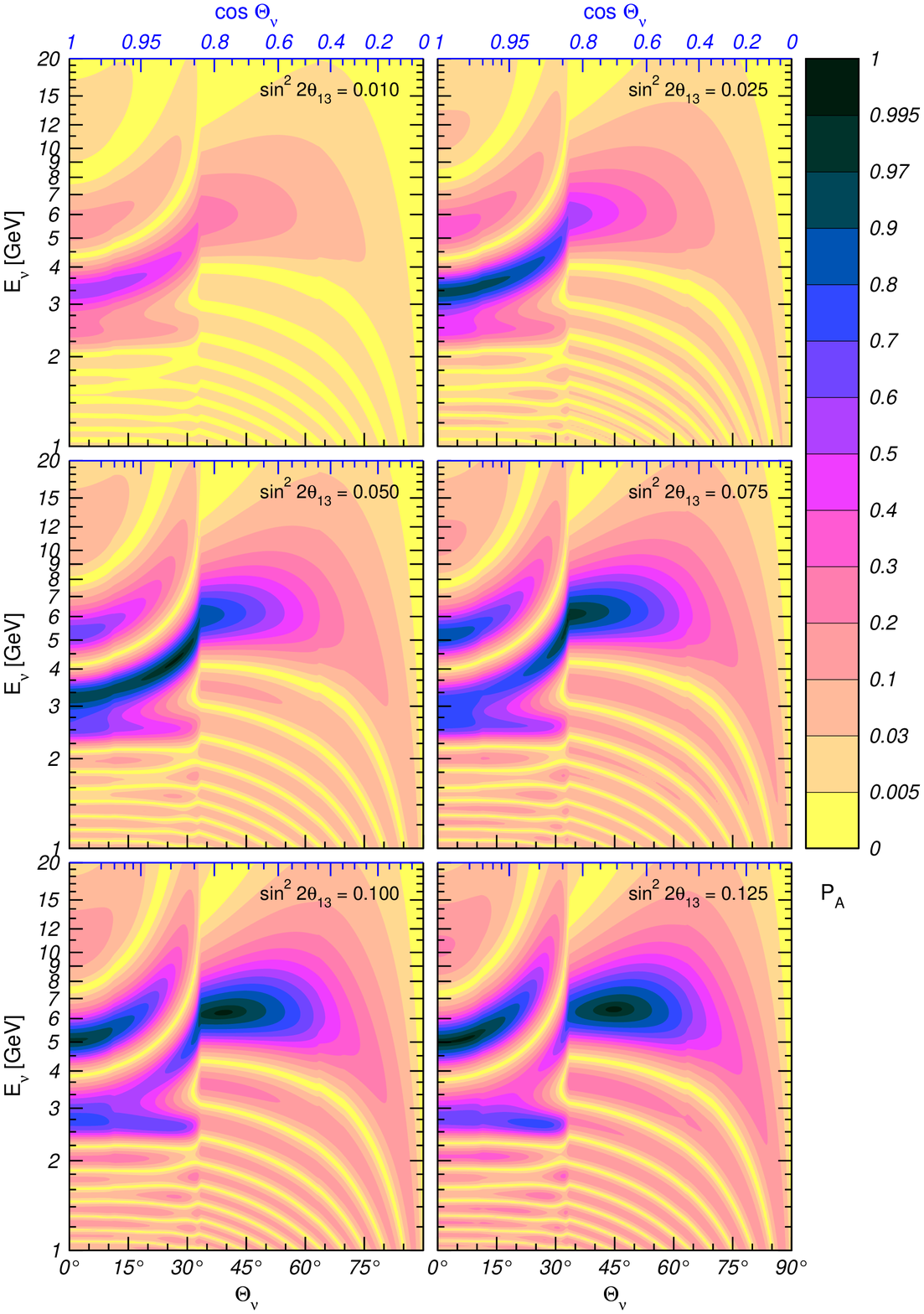}
  \caption{\label{fig:theta13}%
    Neutrino oscillograms of the Earth.  Shown are the contours of
    constant probability $P_A$ (edges of the shadowed regions) in the
    plane of the nadir angle of the neutrino trajectory $\Theta_\nu$ 
    and neutrino energy $E_\nu$, for different values of the mixing 
    angle $\theta_{13}$.}
  }

In Fig.~\ref{fig:theta13} we present the oscillograms for the
transition probability $P_A$ for normal mass hierarchy and several
values of $\sin^2 2\theta_{13}$. The oscillograms have two regions,
separated by $\Theta_\nu = 33.1^\circ$ that corresponds to the nadir
angle of the border between the core and the mantle.
For $\Theta_\nu < 33.1^\circ$, the neutrino trajectories cross the
core of the Earth, so that neutrinos traverse two mantle layers and
one core layer. For brevity we call this part the \emph{core domain}
of the oscillogram. 
Conversely, the region $\Theta_\nu > 33.1^\circ$ corresponds to the
mantle-only crossing trajectories, and we call it the \emph{mantle
domain}.
As follows from the figure, there are several salient, generic
features of the oscillation picture:
\begin{myitemize}
  \item the MSW resonance pattern (resonance enhancement of the
    oscillations) for mantle-only crossing trajectories, with the main
    peak at $E_\nu \sim (5 - 7)$ GeV;
    
  \item three parametric resonance ridges in the core domain, at $E_\nu
    > 3$ GeV;
    
  \item the MSW resonance pattern in the core domain, $E_\nu < 3$ GeV,
    with the core resonance ridge at $E_\nu = 2.5 - 2.7$ GeV;
    
  \item regular oscillatory pattern for low energies: valleys of zero
    probability and ridges in the mantle domain and more complicated
    pattern with local maxima and saddle points in the core domain.
\end{myitemize}

The small windings of the contours at $\Theta_\nu = 65^\circ$ and
$70^\circ$ correspond to the borders of the transition zone between
the inner and upper mantle, while the windings at $\Theta_\nu =
11^\circ$ are due to the border between the inner and outer core. In
what follows we will show that the whole this, apparently complicated,
picture can be understood in terms of different realizations of just
two conditions: the amplitude condition and the phase condition.


\subsection{Constant-density layers approximations}

As we shall see, the main features of the oscillograms can be well
understood using the approximation of constant-density layers for the
density distribution inside the Earth. Furthermore, this consideration
allows us to evaluate the sensitivity of the oscillograms to changes
of the profile. In Figs.~\ref{fig:apx-fixed} and~\ref{fig:apx-path} we
show the oscillograms computed for two different approximations of
this kind:
\begin{myitemize}
  \item Approximation of fixed constant-density layers, when the Earth
    density profile is described by the core and mantle layers of
    constant densities that are the same for all neutrino
    trajectories. This approximation has been widely used in the
    literature. For definiteness, we have taken $\rho_1 = 5.5$
    g/cm$^3$ and $\rho_2 = 11.5$ g/cm$^3$, which roughly correspond to
    the average densities in the core and in the mantle.
    In Fig.~\ref{fig:apx-fixed} we compare the $P_A$-oscillograms
    calculated in this approximation and the exact results. The
    approximation reproduces the oscillation pattern qualitatively
    well, with all the features present. However, quantitatively its
    accuracy is not high above $E_\nu > 3$ GeV, in particular, in the
    resonance region, and for deep trajectories $\Theta_\nu <
    60^\circ$. For instance, one can observe an upward shift of the
    approximate contours compared to the exact ones by about 1 GeV at
    $E_\nu \simeq (6-10)$ GeV. The shift decreases with increasing
    $\Theta_\nu$.
    
  \item Approximation of the path-dependent (trajectory-dependent)
    constant density layers. For each trajectory we find the average
    potentials $\bar{V}_1$ and $\bar{V}_2$ in the mantle and in the
    core, respectively,
    \begin{equation} \label{eq.avpot}
	\bar{V}_i(\Theta_\nu) = \frac{1}{L_i(\Theta_\nu)}
	\int_0^{L_i} V_i(x) \, dx \,,  \qquad i = 1, 2,
    \end{equation}
    where $L_i(\Theta_\nu)$ is the trajectory length in the $i$-th
    layer, and then use the oscillation probability formulas for one
    or three layers of constant densities. In Fig.~\ref{fig:apx-path}
    we show the $P_A$ oscillograms calculated in this approximation.
    One can see that the accuracy of the approximation is noticeably
    better than that of the fixed-density approximation, especially
    for the core-crossing trajectories. Again, the largest difference
    appears for the deep mantle trajectories and $E_\nu > 5$ GeV.
\end{myitemize}

As can be seen from Figs.~\ref{fig:apx-fixed} and~\ref{fig:apx-path},
the accuracy of the constant-density layers approximations improves
with decreasing neutrino energy, the reason being that the matter
effects become less important for small $E_\nu$. These approximations
also in general work better for the core-crossing trajectories because
they are dominated by the evolution in the core and because inside the
core the density changes only by about 30\%, whereas in the mantle it
changes by about a factor of two. For mantle-only crossing
trajectories, the approximations work better for shorter trajectories,
closer to the surface. Here again the matter effects becomes weaker
due to the vacuum mimicking phenomenon~\cite{Wolfenstein:1977ue,
vacmim}.

Thus, the approximations of constant-density layers reproduce
correctly the qualitative structure of the oscillograms obtained with
the realistic PREM density profile of the Earth. All the resonance
peaks, ridges and other structures present for the PREM-profile appear
also in the approximate profile calculations, though their location
and shape is not always well reproduced. Therefore one can use these
approximations for understanding the main qualitative features of the
results for realistic profiles.

\FIGURE[!t]{
  \includegraphics[width=145mm]{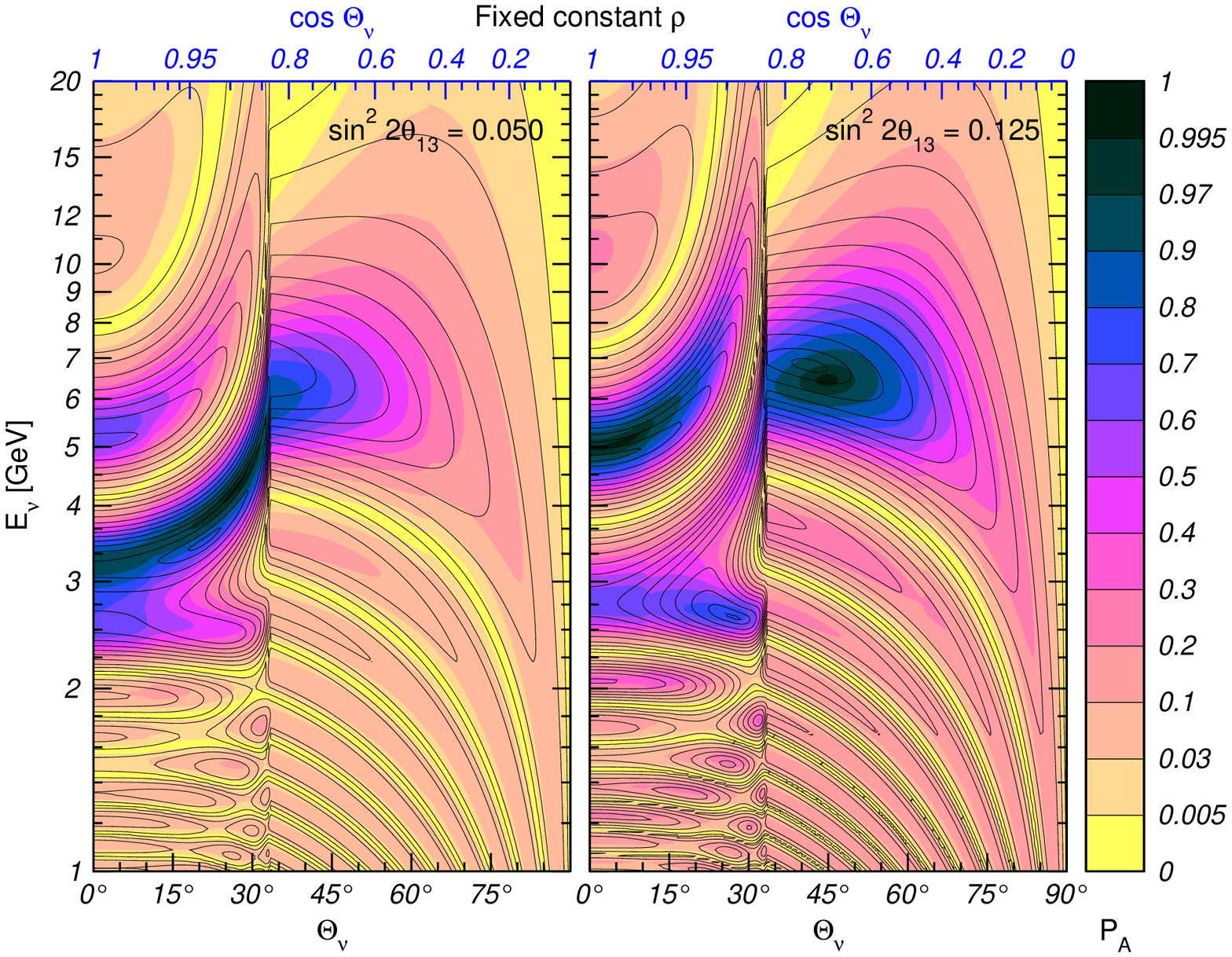}
  \caption{\label{fig:apx-fixed}%
    The $P_A$ oscillograms for the PREM density profile (colored
    regions; grayscale on black-and-white printouts) and for the fixed
    constant-density layers approximation (black curves).}
  }

\FIGURE[!t]{
  \includegraphics[width=145mm]{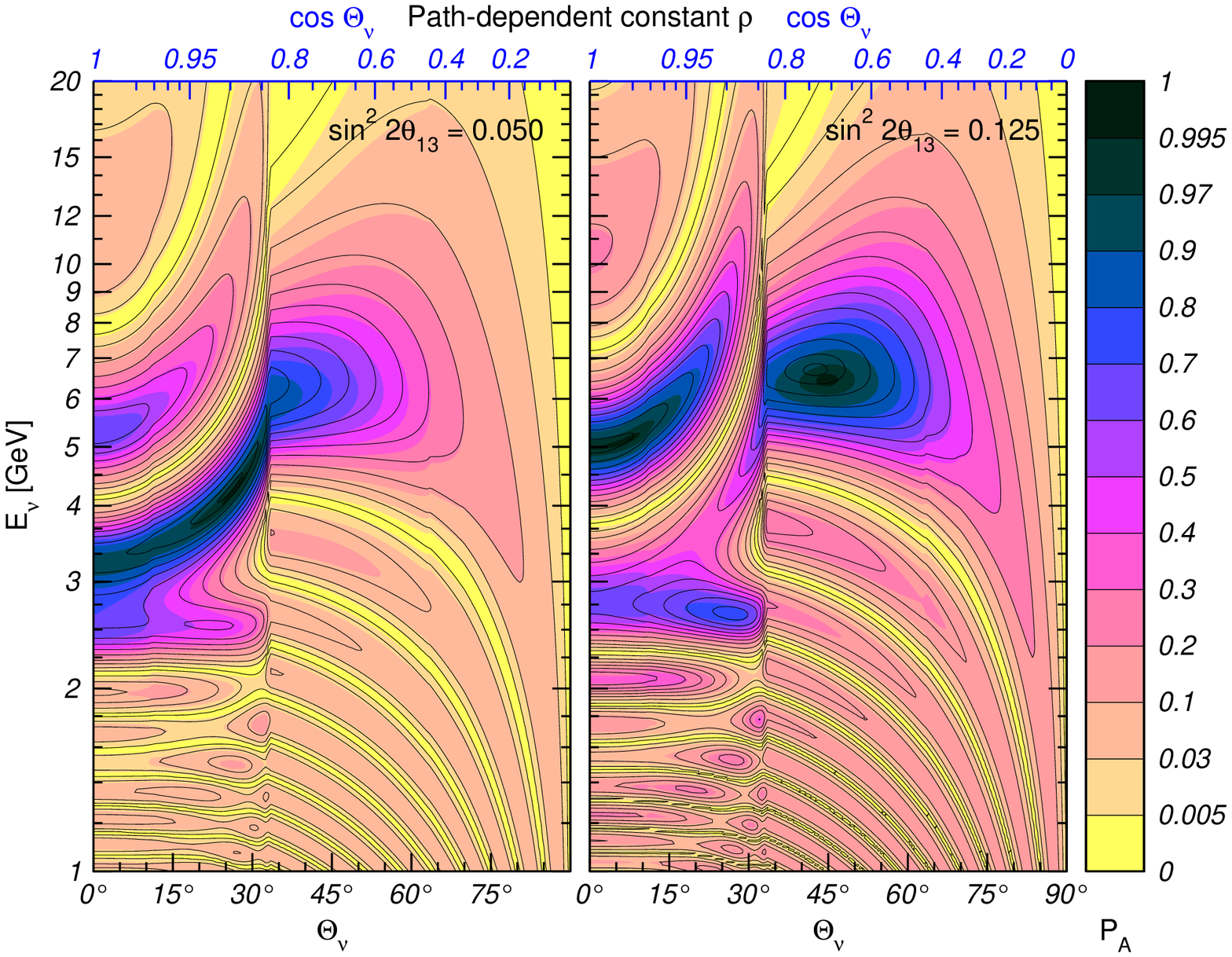}
  \caption{\label{fig:apx-path}%
    The $P_A$ oscillograms for the PREM density profile (colored
    regions; grayscale on black-and-white printouts) and for the
    path-dependent constant-density layers approximation (black
    curves).}
  }


\subsection{Graphical representation}

For illustrative purposes we will also use the graphical
representation of the $2\nu$ oscillations based on their analogy with
spin precession in a magnetic field~\cite{Smirnov:1986ij, graphic}.
Let us introduce the neutrino ``spin'' vector in the flavor space
$\vec{s} = \lbrace s_X ,\, s_Y ,\, s_Z \rbrace$ with the components
\begin{equation}
    s_X(x) = \Re[S_{11}^*(x) \, S_{12}(x)]\,, \quad
    s_Y(x) = \Im[S_{11}^*(x) \, S_{12}(x)]\,, \quad
    s_Z(x) = |S_{11}(x)|^2 - \frac{1}{2}\,.
\end{equation}
These components are essentially the elements of the density matrix.
The evolution equation for the vector $\vec{s}(x)$ can be obtained
from the evolution equation for $S(x)$, given in
Eq.~\eqref{eq:smatreq}:
\begin{equation} \label{eq:evol2}
    \frac{d\vec{s}}{dx} = 2\, \vec{B}(x) \times \vec{s}(x) \,,
\end{equation}
where
\begin{equation} \label{eq:vec-B}
    \vec{B}(x) = \vec{B}(V(x)) = \frac{\Dmq}{4 E_\nu} \LT\lbrace
    \sin2\theta ,\; 0 ,\; \frac{2E_\nu V(x)}{\Dmq} - \cos2\theta
    \RT\rbrace \,.
\end{equation}
In this representation the oscillations in a medium with constant
density is equivalent to a precession of the vector $\vec{s}$ on the
surface of the cone (which we will call the precession cone) with the
axis $\vec{B}(\bar{V})$ and the opening angle $2\theta_m$. According
to Eq.~\eqref{eq:vec-B} the axis of the cone is located in the $(X,\,
Z)$ plane and the angle between the cone axis and the $Z$ axis equals
$2\theta_m$. In a medium with non-constant density profile the cone
axis turns following the change of the angle $2\theta_m(V(x))$. The
opening angle of the cone does not change in the adiabatic case, and
it changes if the adiabaticity is broken.


\section{Physics interpretation of the oscillograms}
\label{sec:interposc}

In this section we give a physics interpretation and description of
the four main structures of the oscillograms mentioned in
Sec.~\ref{sec:oscillogr}, as well as of the contours of zero
probability.


\subsection{Resonance enhancement of oscillations in the mantle}

The oscillation pattern in the mantle is determined by the resonance
enhancement of the oscillations~\cite{Wolfenstein:1977ue,
Mikheev:1986gs}. In the constant-density approximation, for a given
$\Theta_\nu$ the probability $P_A$ is an oscillatory function of
energy, which is inscribed in the resonance curve $\sin^2 2
\theta_m(E_\nu)$. The position of the maximum of the resonance peak is
given by the MSW resonance condition:
\begin{equation} \label{eq:Eres}
    E_\nu = E_R (\Theta_\nu) =
    \frac{\Dmq_{31} \cos 2 \theta_{13}}
    {2\bar{V}_1(\Theta_\nu)} \,,
\end{equation}
where $\bar{V}_1(\Theta_\nu)$ is the average value of the potential
along the trajectory characterized by $\Theta_\nu$ (see
Eq.~\eqref{eq.avpot}). Eq.~\eqref{eq:Eres} determines the resonance
line in the $(E_\nu,\, \Theta_\nu)$ plane. With the increase of
$\Theta_\nu$ the average potential decreases and consequently $E_R$
increases. According to Fig.~\ref{fig:theta13}, for the MSW resonance
in the mantle we have $E_R \sim 6$ GeV. The resonance width $\Delta
E_\nu / E_R \sim 2\tan 2\theta_{13}$. The condition $E_\nu =
E_R(\Theta_\nu)$ ensures that the amplitude of oscillations is maximal
and we will call it the \emph{amplitude} (or \emph{resonance})
\emph{condition}.

Another condition that should be met to obtain the absolute maximum of
the transition probability, $P_A = 1$, is the \emph{phase condition}:
\begin{equation} \label{eq:Pline}
    2\phi(E_\nu, \Theta_\nu) =
    2\omega(\bar{V}, E_\nu) L(\Theta_\nu) = (2k+1)\pi
\end{equation}
which means that the oscillation phase should be an odd integer of
$\pi$. Since the size of the Earth is comparable with the refraction
length in the mantle $l_0 = 2\pi / \bar{V}$, the condition of the
absolute maximum~\eqref{eq:Pline} can only be fulfilled for $k=0$,
\ie\ when the length of the neutrino trajectory coincides with the
half of the oscillation length in matter. The phase
condition~\eqref{eq:Pline} gives another curve in the $(E_\nu ,\,
\Theta_\nu)$ plane. The intersection of the resonance
line~\eqref{eq:Eres} and the phase condition line~\eqref{eq:Pline}
gives the absolute maximum of $P_A$. Combining~\eqref{eq:Eres}
and~\eqref{eq:Pline} we conclude that the absolute maximum corresponds
to the situation when for the resonance energy the oscillation phase
is $\pi$:
\begin{equation} \label{eq:phasecond}
    2\omega(\bar{V}, E_R) L(\Theta_\nu) = \pi\,.
\end{equation}
Since $L=2R \cos\Theta_\nu$ and at the resonance $2\omega = \sin
2\theta_{13} (\Dmq_{31} /2E_\nu)$, the phase
condition~\eqref{eq:phasecond} yields
\begin{equation} \label{eq:phase2}
    \cos\Theta_\nu = \frac{\pi E_\nu}{R \sin 2\theta_{13}\Dmq_{31}}\,.
\end{equation}
With the increase of $\theta_{13}$ the peak shifts to smaller values
of $\cos\Theta_\nu$ (see Fig.~\ref{fig:theta13}).

Let us now reformulate the resonance and the phase conditions for the
case of varying density within the mantle. For this we first express
these conditions in terms of the elements of the evolution matrix
using the explicit results for matter of constant density,
Eqs.~\eqref{eq:U1} and~\eqref{eq:Ubar}, and then use the obtained
conditions also in the case of varying matter density.

As concerns to the resonance condition, $\cos 2 \theta_m = 0$, it can
be written according to Eqs.~\eqref{eq:U1} and~\eqref{eq:ab1} as
$\alpha = \alpha^*$, \ie\
\begin{equation} \label{eq:res1layer}
    S_{11}^{(1)} = S_{22}^{(1)} \,,
\end{equation}
or equivalently,
\begin{equation} \label{eq:res1layera}
    \Im S_{11}^{(1)} = 0 \,,
\end{equation}
where the superscript indicates the number of layers. This
generalization, however, goes beyond the original MSW-resonance
condition. For the constant density case it gives
\begin{equation} \label{eq:const-res}
    \cos 2 \theta_m \sin\phi = 0\,.
\end{equation}
This condition has two realizations: the original one
\begin{equation} \label{eq:g-con1}
    \cos 2 \theta_m = 0\,,
\end{equation}
and
\begin{equation} \label{eq:g-con2}
    \phi = \pi k, \qquad(k = 1, 2, \ldots)
\end{equation}
and the latter corresponds to $P_A = 0$, which is realized at low
energies.
In the case of varying density (\eg, in the mantle of the Earth),
there is no factorization of the resonance
condition~\eqref{eq:res1layera} into path-dependent and
path-independent factors, as in Eq.~\eqref{eq:const-res}. However,
when the density varies slowly enough along the neutrino path, in
certain limits Eq.~\eqref{eq:res1layera} is still approximately
realized in the form of Eqs.~\eqref{eq:g-con1} or~\eqref{eq:g-con2}.
At low energies the MSW resonance condition is not fulfilled in the
Earth's mantle, and therefore the only possible realization
of~\eqref{eq:res1layera} is the one in Eq.~\eqref{eq:g-con2}.
In contrast to this, for large energies and small distances
Eq.~\eqref{eq:g-con2} is not satisfied, and the only way
Eq.~\eqref{eq:res1layera} can be implemented is through the MSW
resonance~\eqref{eq:g-con1}. In the intermediate region neither of the
two conditions~\eqref{eq:g-con1} and~\eqref{eq:g-con2} is satisfied.
Thus, in the resonance region Eq.~\eqref{eq:res1layera} interpolates
between the MSW resonance condition and the condition $\phi=\pi$.

As for the phase condition, $\phi = \pi/2 + \pi k$, again we can
rewrite it in terms of the elements of the evolution matrix, and from
Eqs.~\eqref{eq:U1} and~\eqref{eq:ab1} we get
\begin{equation} \label{eq:pha1layera}
    \Re\alpha \equiv \Re S_{11}^{(1)} = 0 \,.
\end{equation}
The absolute maximum of the transition probability occurs when
\emph{both} conditions~\eqref{eq:res1layera} and~\eqref{eq:pha1layera}
are satisfied simultaneously. In this case $S_{11}^{(1)}=0$ and $P_A =
1$. This situation corresponds to $\phi = \pi/2$, and from
Fig.~\ref{fig:theta13} we see that it is realized only for $\sin^2
2\theta_{13} \gtrsim 0.08$.


\subsection{Parametric ridges and generalized resonance condition}

For core-crossing trajectories and $E_\nu > 3$ GeV the oscillatory
picture is characterized by three ridges of enhanced oscillation
probability. The ridges are the curves along which the probability
decreases most slowly from its local or absolute maximum. The
lower-energy ridge, which we will call the ridge ``$A$'', corresponds
to the energies in between those of the MSW resonances in the core and
in the mantle, $E_\nu > (3 - 6)$ GeV. It was interpreted in~\cite{LS,
Akhm2, ADLS} as being due to the parametric enhancement of neutrino
oscillations. The other two ridges, ``$B$'' and ``$C$'', extend above
the MSW resonance energy in the mantle. They were also identified as
the effects of the parametric enhancement of neutrino
oscillations~\cite{AMS1}. Here we shall further elaborate on this
interpretation.

Recall that the parametric resonance occurs in oscillating systems
with varying parameters when the rate of the parameter change is in a
special correlation with the values of the parameters themselves.
Neutrino oscillations in matter can undergo parametric enhancement if
the length and size of the density modulation are correlated in a
certain way with neutrino parameters~\cite{ETC,Akhm1}. This
enhancement is completely different from the MSW effect; in
particular, no level crossing is required.

An example admitting exact analytic solution, and in fact, relevant
for our discussion, is the ``castle wall'' density
profile~\cite{Akhm1,Akhm2}. This is a periodic step function with one
period consisting of two layers of widths $L_1$ and $L_2$ and electron
number densities $N_1$ and $N_2$. For 2-flavor neutrino oscillations,
the evolution matrix over one period of density modulation can be
written as
\begin{equation}
    \label{eq:U2}
    S^{(2)} = Y \mathbbm{1} - i {\boldsymbol\sigma} \cdot
    \mathbf{X} \,,
\end{equation}
where $Y$ and $\mathbf{X} = (X_1,\, X_2,\, X_3)$ are real parameters
satisfying $Y^2 + \mathbf{X}^2 = 1$ and $\sigma_i$ are the Pauli
matrices in the flavor space. The oscillation probability for an
arbitrary number of layers traversed by neutrinos can be written as a
product of the amplitude
\begin{equation}
    A=\frac{X_1^2 + X_2^2}{X_1^2 + X_2^2 +  X_3^2} \,,
\end{equation}
which does not depend on the number of the layers, and an oscillating
factor, which depends on this number~\cite{Akhm2}. The amplitude $A$
reaches its maximum when $X_3=0$, or explicitly~\cite{Akhm2}
\begin{equation} \label{eq:parcond}
    X_3 = - (s_1 c_2 \cos 2 \theta_1  + s_2 c_1 \cos 2\theta_2) = 0,
\end{equation}
where $s_{1,2} \equiv \sin \phi_{1,2}$ and $c_{1,2} \equiv
\cos\phi_{1,2}$. Here $2\phi_{1,2}$ are the oscillation phases
acquired in the layers $1$ and $2$, and $\theta_{1,2}$ are the
corresponding mixing angles in matter. Note that in the
constant-density limit ($N_1 = N_2$ or $\theta_1 = \theta_2 =
\theta_m$) this condition reduces to Eq.~\eqref{eq:const-res} with
$\phi = \phi_1 + \phi_2$. For $\sin \phi \neq 1$ it coincides with the
MSW resonance condition, which is the maximum amplitude condition for
oscillations in a matter of constant density.

As was pointed out above, the Earth density profile seen by neutrinos
with core-crossing trajectories can be very well approximated by three
layers of constant densities, which is nothing but a piece of the
castle wall profile; the layers ``1'' and ``2'' have to be identified
with the Earth's mantle and core, respectively. It has been
demonstrated that the parametric resonance
condition~\eqref{eq:parcond} can be satisfied for oscillations of
core-crossing neutrinos for a rather wide range of nadir angles both
at intermediate energies~\cite{LS,P,Akhm2} and high
energies~\cite{AMS1}, leading to significant enhancement of the
oscillation probability. Thus, despite the fact that the Earth's
density profile consists of only three layers, the effects can be
quite substantial.
This show the relevance of the parametric resonance interpretation of
this phenomenon, even for just ``1.5 periods'' of density
modulation~\cite{Akhmedov:1999va}.

The parametric resonance condition~\eqref{eq:parcond} can be readily
generalized to the case of non-constant densities in the mantle and
the core of the Earth, though the generalization is not unique.
Indeed, according to~\eqref{eq:U2} the condition $X_3 = 0$ can be
written in terms of elements of the evolution matrix for two layers as
the equality of the diagonal elements:
\begin{equation} \label{eq:genres}
    S_{11}^{(2)} = S_{22}^{(2)}.
\end{equation}
We will use this equality for an arbitrary density distribution within
the layers and call it \emph{the generalized resonance condition}.
Note that parameterization~\eqref{eq:U2} of the matrix $S^{(2)}$
implies that the generalized resonance condition can also be
formulated as the requirement that the diagonal element of $S^{(2)}$
be real: $\Im S_{11}^{(2)} = 0$.

In the right panels of Fig.~\ref{fig:collinear} we show the curves of
the generalized parametric resonance condition. Apparently, all the
three parametric ridges are very well described by these curves.

Since condition~\eqref{eq:parcond} does not depend on the number of
layers traversed by neutrinos, it only ensures that the parametric
oscillations occur with the maximal depth, that is, after some number
of periods of density modulation $P \approx 1$. In general, the
condition of maximal transition probability for a given number of
layers need not coincide with the condition of maximal amplitude of
the parametric oscillations. Thus, the fact that
Eq.~\eqref{eq:parcond} and the generalized condition~\eqref{eq:genres}
give the correct description of ridges for the Earth density profile
with three layers is rather non-trivial.


\subsection{Parametric effects and collinearity condition}
\label{sec:nonconst}

Another approach to the generalization of the parametric resonance
condition to the case of the layers of non-constant density is based
on the consideration of the evolution amplitudes in the individual
layers. For the particular case of two layers of constant densities,
similar considerations have been presented in~\cite{CP}.

For density profiles consisting of two layers we have
\begin{equation} \label{eq:U21}
    S^{(2)} =  S_2 \, S_1 =
    \begin{pmatrix}
	S_{11}^{(2)} & S_{12}^{(2)} \\
	-S_{12}^{(2)*} & S_{11}^{(2)*}
    \end{pmatrix} \,,
\end{equation}
where
\begin{equation} \label{eq:a12b21}
    S_{11}^{(2)} = \alpha_2 \, \alpha_1 - \beta_2 \, \beta_1^* \,,
    \qquad
    S_{12}^{(2)} = \alpha_2 \, \beta_1 + \beta_2 \, \alpha_1^* \,,
\end{equation}
and $\alpha_i$, $\beta_i$ for each layer have been defined in
Eq.~\eqref{eq:ab1}. The sum of the two complex numbers in the
transition amplitude $S_{12}^{(2)}$ can potentially lead to the
largest possible result (if they add in the same phase and not in the
anti-phase) if the two contributions to $S_{12}$ have the same complex
phase (modulo $\pi$):
\begin{equation} \label{eq:collin}
    \arg(\alpha_2 \beta_1) = \arg(\beta_2 \alpha_1^*)
    \quad \mbox{mod}(\pi) \,.
\end{equation}
It can also be rewritten as
\begin{equation} \label{eq:collin2}
    \arg(\alpha_1 \alpha_2 \beta_1) = \arg(\beta_2)
    \quad \mbox{mod}(\pi) \,.
\end{equation}
We shall call this condition the \emph{collinearity condition}. It is
an extremality condition for the two-layer transition probability
under the constraint of fixed transition probabilities in the
individual layers. In other words, if the absolute values $|\beta_i|$
of the transition amplitudes are fixed while their arguments are
allowed to vary, the transition probability reaches an extremum when
these arguments satisfy~\eqref{eq:collin} or~\eqref{eq:collin2}. For a
realistic situation (neutrino oscillations in the Earth), changes of
$E_\nu$ and $\Theta_\nu$ produce correlated changes of the arguments
and the absolute values of the individual amplitudes, and therefore in
general the condition~\eqref{eq:collin} may not correspond to extrema
precisely.

If the layers 1 and 2 have constant densities, then $\alpha_i=c_i+i
\cos 2\theta_i\,s_i$, $\beta_i=-i\sin 2\theta_i\,s_i\,$
(see~\eqref{eq:ab1}), and the condition in Eq.~\eqref{eq:collin}
reproduces the parametric resonance condition~\eqref{eq:parcond}, \ie\
$X_3=0$.  Under this condition the diagonal elements of $S^{(2)}$ are
real; therefore, in the case of constant-density layers the
collinearity condition, the generalized resonance condition and the
parametric resonance condition coincide.

Denoting by $\chi_{\alpha i}$ and $\chi_{\beta i}$ the arguments of
the complex amplitudes $\alpha_i$ and $\beta_i$, respectively, one can
rewrite the collinearity condition as
\begin{equation} \label{eq:coll}
    \chi_{\alpha 1} + \chi_{\alpha 2} =
    \chi_{\beta 2} - \chi_{\beta 1} \quad \mbox{mod}(\pi)\,.
\end{equation}

Consider now the case of three layers of in general varying densities.
For the elements of the evolution matrix $S^{(3)}$ one obtains
\begin{align}
    \label{eq:def-a}
    S_{11}^{(3)} &= \alpha_3\, S_{11}^{(2)} - \beta_3\, S_{12}^{(2)*}
    = \alpha_3\, \alpha_2\, \alpha_1 - \alpha_3\, \beta_2\, \beta_1^*
    - \beta_3\, \alpha_2^*\, \beta_1^*
    - \beta_3\, \beta_2^*\, \alpha_1 \,,
    \\
    \label{eq:def-b}
    S_{12}^{(3)} &= \alpha_3\, S_{12}^{(2)} + \beta_3\, S_{11}^{(2)*}
    = \alpha_3\, \alpha_2\, \beta_1 + \alpha_3\, \beta_2\, \alpha_1^*
    + \beta_3\, \alpha_2^*\, \alpha_1^*
    - \beta_3\, \beta_2^*\, \beta_1 \,.
\end{align}
In the case of neutrino oscillations in the Earth, the third layer is
just the second mantle layer, and its density profile is the reverse
of that of the first layer. The evolution matrix for the third layer
is therefore the transpose of that for the first one~\cite{Tviol},
\ie\ $\alpha_3=\alpha_1$, $\beta_3=-\beta_1^*$, and the expression for
$S_{12}^{(3)}$ can be written as
\begin{equation} \label{eq:res2}
    S_{12}^{(3)} = \alpha_1 \alpha_2 \beta_1
    - \alpha_1^* \alpha_2^* \beta_1^*
    + |\alpha_1|^2 \beta_2 + |\beta_1|^2 \beta_2^* \,.
\end{equation}
Note that $\beta_2$ is pure imaginary because the core density profile
is symmetric. Therefore the amplitude $S_{12}^{(3)}$ in
Eq.~\eqref{eq:res2} is also pure imaginary, as it must be because the
overall density profile of the Earth is symmetric as well. It is easy
to see that if the collinearity condition for two
layers~\eqref{eq:collin2} is satisfied, then not only the full
amplitude $S_{12}^{(3)}$, but also \emph{each} of the four terms on
the right hand side of Eq.~\eqref{eq:res2} is pure imaginary. We
therefore conclude that if the collinearity condition is satisfied for
two layers, then it is automatically satisfied for three layers as
well. It should be stressed that this is a consequence of the facts
that the density profile of the third layer is the reverse of that of
the first layer and that the second layer has a symmetric profile.
Once again, the collinearity of all the contributions to the
transition amplitude potentially leads to the maximal total transition
probability for given transition probabilities in each layer.

Let us now confront the generalized resonance condition and the
collinearity condition. For two layers the generalized resonance
condition can be written as
\begin{equation} \label{eq:res2a}
    -X_3 = \Im S_{11}^{(2)} =
    \Im(\alpha_2 \alpha_1 - \beta_2 \beta^*_1) = 0
\end{equation}
or, in terms of the moduli and complex phases of the amplitudes,
\begin{equation} \label{eq:res3}
    |\alpha_2 \alpha_1| \sin (\chi_{\alpha 1} + \chi_{\alpha 2}) =
    |\beta_2 \beta_1| \sin ( \chi_{\beta 2} - \chi_{\beta 1})\,.
\end{equation}
Comparing~\eqref{eq:res3} and~\eqref{eq:coll} we come to the following
conclusions:

\begin{myitemize}
  \item If the generalized resonance condition~\eqref{eq:res3} has to
    be satisfied independently of the moduli,\footnote{The same
    condition is required for the collinearity condition to give
    extrema.} then $\chi_{\alpha 1} + \chi_{\alpha 2} = \chi_{\beta 1}
    - \chi_{\beta 2} = \pi k$, and the collinearity
    condition~\eqref{eq:coll} is satisfied too.
    
  \item If the collinearity condition is fulfilled, then the
    generalized resonance condition~\eqref{eq:res3} can be rewritten
    as
    \begin{equation}
	\sin(\chi_{\beta 1} - \chi_{\beta 2})
	\LT[ |\alpha_2 \alpha_1|\pm |\beta_1 \beta_2| \RT] = 0\,,
    \end{equation}
    which is satisfied provided that
    \begin{equation} \label{eq:cond}
	\chi_{\beta 1} = \chi_{\beta 2} \quad \mbox{mod}(\pi)\,.
    \end{equation}
    This condition is, in particular, fulfilled when the density
    profiles in both layers are symmetric (\eg, in the
    constant-density layers case), since in that case $\chi_{\beta 1}
    = \chi_{\beta 2} =\pm \pi/2$ as a consequence of T-invariance.
\end{myitemize}

Thus, we conclude that for symmetric profiles in each of the two
layers (and, in particular, for constant-density layers) the
generalized resonance condition and the collinearity condition
coincide. For layers of non-symmetric densities, the two conditions
differ. As can be seen in the right panels of
Fig.~\ref{fig:collinear}, both conditions describe the parametric
enhancement ridges in the oscillograms quite well. This is a
consequence of the fact that the matter density profiles felt by
neutrinos traversing the Earth can be well approximated by
path-dependent constant density layers.


\subsection{Extrema and saddle points}

As has been discussed above, in the constant density case the
positions of maxima of the oscillation probability are determined by
two conditions: the amplitude condition and the phase condition.

For matter of non-constant density, the collinearity condition or the
2-layer generalized resonance condition can be considered as the
generalizations of the \emph{amplitude condition}. Taking into account
that the density profile of the Earth's core is symmetric (\ie\
$\Re\beta_2 = 0$), one can rewrite the collinearity
condition~\eqref{eq:collin2} as
\begin{equation} \label{eq:collin3}
    \Re(\alpha_1 \alpha_2 \beta_1) = 0\,.
\end{equation}
Compared with Eq.~\eqref{eq:collin2}, this condition has the practical
advantage of not being trivially satisfied for $\beta_2=0$, which has
no physically relevant implications. For trajectories crossing only
the mantle of the Earth, one has to set $\alpha_2=1$ in
Eq.~\eqref{eq:collin3} and the collinearity condition reduces to
$\Re(\alpha_1 \beta_1) = 0$. Recall that in the limit of constant
matter density in the mantle this latter condition reduces to the
condition of maximal mixing in matter (barring $\phi_1 = \pi k$).

The phase condition $\phi = \pi/2 + \pi k$ of the constant-density
case can be generalized for varying density by expressing it in terms
of the elements of the evolution matrix. According to~\eqref{eq:ab1},
for one layer and the phase $\phi=\pi/2 +\pi k$ the product of
amplitudes $\alpha \beta^* = \pm \sin 2\theta \cos 2\theta$ is real.
Therefore, we can generalize the phase condition $\Im(\alpha \beta^*)$
taking instead of $\alpha$ and $\beta$ the elements of the evolution
matrix (amplitudes) for an arbitrary profile:
\begin{equation} \label{eq:phase1}
    \Im(S_{11} S_{12}^*)  = 0.
\end{equation}
Furthermore, since for symmetric density profiles $S_{12}$ is pure
imaginary, Eq.~\eqref{eq:phase1} gives
\begin{equation} \label{eq:phase2a}
    \Re S_{11} = 0\,,
\end{equation}
\ie, the phase condition is fulfilled when $S_{11}$ is pure imaginary.

In the graphical representation of neutrino oscillations based on
their analogy with spin precession in a magnetic field (see
Sec.~\ref{sec:graphical}), Eq.~\eqref{eq:phase1} corresponds to the
condition that the neutrino ``spin'' vector is in the $(s_X ,\, s_Z)$
plane. As can be easily seen, this is equivalent to the requirement
that the transition probability be stationary with respect to small
variations of the total distance $L$ traveled by neutrinos:
$dP_A/dL=0$.

In the right panels of Fig.~\ref{fig:collinear} we plot the
collinearity (amplitude) condition, the generalized resonance
condition and the phase condition for two different values of $\sin^2
2\theta_{13}$.

\PAGEFIGURE{
  \includegraphics[width=140mm]{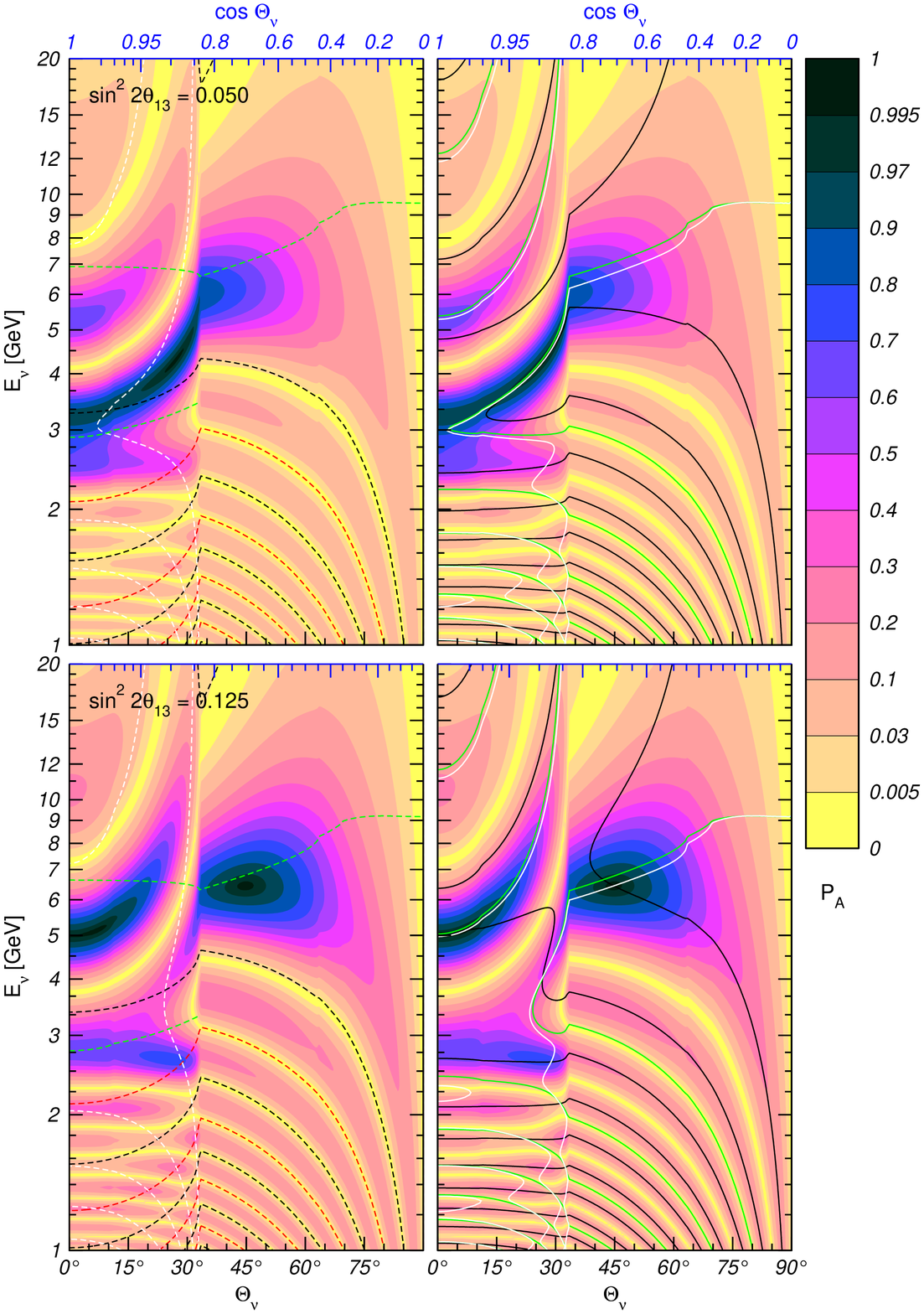}
  \caption{\label{fig:collinear}%
    $P_A$ oscillograms and curves of phase and amplitude conditions
    for the PREM profile. Left panels: the MSW resonance (dashed
    green), $\Re\alpha_1 = 0$ (dashed black), $\Im\beta_1 = 0$ (dashed
    red) and $\Re\alpha_2 = 0$ (dashed white). Right panels:
    collinearity condition (solid white), generalized resonance
    condition (solid green) and phase condition (solid black).}
  }

As follows from the figure, the simultaneous fulfillment of the phase
and amplitude conditions leads not only to absolute maxima of the
transition probability ($P_A = 1$), as in the constant density case,
but also to local maxima and saddle points. This is an effect of the
multi-layer medium.  To figure out why this happens, we will use the 
explicit results for three layers of constant densities. In this
case~\cite{Akhm2}
\begin{equation} \label{eq:U11}
    S_{11}^{(3)} = Z - i W_3 \,,
\end{equation}
where
\begin{equation} \label{eq:Z}
    Z = 2 c_1 Y - c_2\,, \qquad
    W_3 = -(2s_1 Y \cos 2\theta_1 + s_2 \cos 2 \theta_2) \,,
\end{equation}
and
\begin{equation} \label{eq:Y}
    Y = c_1 c_2 -  s_1 s_2 \cos 2(\theta_1 - \theta_2) \,.
\end{equation}
Both $Z$ and $W_3$ are real and therefore the phase
condition~\eqref{eq:phase2a} gives $Z = 0$, \ie:
\begin{equation} \label{eq:z0cond}
    2 c_1 Y - c_2 = 0 \qquad \text{(phase condition)}.
\end{equation}
Also in this case
\begin{equation} \label{eq:probphase}
    P_A = 1 - W_3^2.
\end{equation}
The collinearity condition~\eqref{eq:collin3} reduces to $\sin
2\theta_1 s_1 X_3 = 0\,$ for constant density layers, and since $\sin
2\theta_1 \ne 0$, we have
\begin{equation} \label{eq:collin4}
    s_1 X_3 = 0 \qquad \text{(amplitude condition)}.
\end{equation}

Let us analyze possible realizations of these two conditions.
According to~\eqref{eq:z0cond}, there are two ways to satisfy the
phase condition: (i) $c_1 \neq 0$, $Y = c_2/2c_1$ and (ii) $c_1 = 0$,
$c_2 = 0$. The amplitude condition~\eqref{eq:collin4} also has two
realizations: (i) $X_3 = 0$, ($s_1 \neq 0$), and (ii) $s_1 = 0$, ($X_3
\neq 0$). As we will see below, different consistent combinations of
these realizations lead to absolute maxima, local maxima or saddle 
points of $P_A$.  

In the non-constant density case, we will use Eq.~\eqref{eq:collin3}
as the general amplitude condition and Eq.~\eqref{eq:phase2a} as the
general phase condition. Using Eq.~\eqref{eq:ab1}, we generalize the
other equalities discussed above as
\begin{align}
    \label{eq:otherc1}
    c_1 = 0 ~ & \rightarrow~ \Re\alpha_1 = 0,
    \\
    \label{eq:otherc2}
    c_2 = 0 ~ & \rightarrow~ \Re\alpha_2 = 0,
    \\
    \label{eq:others1}
    s_1 = 0 ~ & \rightarrow~ \Im\beta_1 = 0.
\end{align}
In the left panels of Fig.~\ref{fig:collinear} we show the curves that
correspond to different conditions. From the constant-density limit it
is clear why the curves $\Im\beta_1 = 0$ always coincide with some of
the amplitude condition curves: this follows from the fact that
$\Im\beta_1 = 0$ is a particular solution of the amplitude condition. 

Let us now consider all consistent realizations of the amplitude and
phase conditions, using the terminology of the constant density
approximation:

\textitem{$\bullet$} $X_3 = 0$ ($s_1 \neq 0$) (amplitude); $Y = c_2 /
2c_1$ ($c_1 \neq 0$) (phase). 
Plugging this expression for the phase condition in~\eqref{eq:Z}, we
obtain
\begin{equation} \label{eq:W3}
    W_3 = \frac{X_3}{c_1} \,.
\end{equation}
Since $X_3 = 0$ by the amplitude condition, we have $W_3 = 0$ and, 
consequently, from Eq.~\eqref{eq:probphase} $P_A = 1$. Thus, in the
constant density layers approximation a simultaneous fulfillment of
the phase and collinearity conditions should lead to $P_A = 1$,
provided that $c_1 \neq 0$ or $\pm 1$.
This possibility is realized only on the ridge $A$, at a point where
the two curves that correspond to the generalized amplitude and phase
conditions cross. The other curves, depicting the
conditions~\eqref{eq:otherc1}, \eqref{eq:otherc2}
and~\eqref{eq:others1}, cannot pass through this point. Notice that at
this crossing point the oscillation half-phases in the core and mantle
differ from $\pi/2$ and $\pi$, and there is a nontrivial interplay
between the phases and mixing angles in the parametric resonance 
condition~\eqref{eq:parcond}.

\textitem{$\bullet$} $c_1 = 0$, $c_2 = 0$ (the latter equality follows
from Eq.~\eqref{eq:z0cond}) (the phase condition).
In this case the amplitude condition, $X_3 = 0$, is satisfied
automatically. Using the explicit formulas for $Y$~\eqref{eq:Y},
$W_3$~\eqref{eq:Z} and Eq.~\eqref{eq:probphase}, we obtain $Y = \pm
\cos 2(\theta_1 - 2\theta_2)$ and
\begin{equation} \label{eq:saddle}
    P_A =\sin^2 (4\theta_1 - 2\theta_2) \,.
\end{equation}
This case corresponds to the core and mantle half-phases equal to
$\pi/2 + \pi k$. It has a simple graphical representation, when it is
enough to consider neutrino ``spin'' vector in the $(X ,\, Z)$ plane
(see Sec.~\ref{sec:graphical} below). This representation immediately
leads to the expression in Eq.~\eqref{eq:saddle} for the transition
probability \cite{LS}, and shows that this probability has a maximum
for neutrino energies between those of the core and mantle MSW
resonances (where \mbox{$\cos 2 \theta_1 > 0$} and $\cos 2\theta_2 <
0$), provided that $(\theta_2 - \theta_1) > \pi/4$, and above the MSW
resonances, where \mbox{$\cos 2\theta_{1,2} < 0$}. Below the
resonances and between the resonances for \mbox{$(\theta_2 -
\theta_1)$} $\leq \pi/4$, Eq.~\eqref{eq:saddle} corresponds to a
saddle point of the transition probability. This agrees with the
findings in~\cite{Akhm2}.

As follows from Fig.~\ref{fig:collinear}, for $E_\nu < 2.5$ GeV (below
the MSW resonance energies) the intersections of the curves
$\Re\alpha_1 = 0$ and $\Re\alpha_2 = 0$ (the analogues of $c_1 = c_2 =
0$ in the case of non-constant densities within the layers) mark the
positions of the saddle points. Also in these points the curves that
correspond to the general amplitude and phase conditions intersect
(since both conditions are fulfilled). At high energies the condition
$c_1 = 0$ is not satisfied: the phase in the mantle is below $\pi$.
Therefore, the maxima of the transition probability are not achieved. 

\textitem{$\bullet$} $s_1 = 0$ ($c_1=\pm 1$) (the amplitude
condition), $Y = 2 c_2/c_1$ (the phase condition).
The latter equality can be written as $Y = \pm 2 c_2$. On the other
hand, from the explicit expression for $Y$ (Eq.~\eqref{eq:Y}) and for
$s_1 = 0$ one has $Y = \pm c_2$. Obviously, the two expressions for
$Y$ are consistent only if $c_2 = 0$. For $s_1 = c_2 = 0$ we find
from~\eqref{eq:Z} $W_3 = \pm \cos 2\theta_2$, and consequently,
\begin{equation} \label{eq:matterpeak}
    P_A = \sin^2 2\theta_2.
\end{equation}
This realization corresponds to the oscillation half-phase in the
mantle equal to $\pi$ and therefore to the absence of the oscillation
effect in the mantle (in the approximation of constant-density
matter). The whole oscillation effect is then due to the evolution in
the core. The oscillation half-phase in the core is a semi-integer of
$\pi$ ($c_2 = 0$).  Eq.~\eqref{eq:matterpeak} thus simply corresponds
to the maximum oscillation probability for a given mixing angle in
matter $\theta_2$. At the intersection points of the curves depicting
the conditions $s_1 = 0$ and $c_2 = 0$ the probability takes values
that correspond to the resonance enhancement of the oscillations in
the core. For non-constant density, according to 
Fig.~\ref{fig:collinear} (left panels), the intersections of the
curves $\Im\beta_1 = 0$ and $\Re\alpha_2 = 0$ (which are analogues of
$s_1 = 0$ and $c_2 = 0$) correspond to local maxima. These points lie
at energies below 2.5 GeV. For higher energies, due to the large
oscillation lengths, the oscillation half-phase in the mantle is
smaller than $\pi$ and the condition $s_1 = 0$ is not fulfilled.

\textitem{}Notice that along the lines $\Im\beta_1 = 0$ (or $\phi_1
\approx \pi k$) the oscillation effects correspond to the resonance
enhancement in the core, whereas the saddle points are situated along
the lines $\Re\alpha_1 = 0$ (or $\phi_1 \approx \pi/2 + \pi k$).


\subsection{Absolute minima and maxima of the transition probability}

As follows from Fig.~\ref{fig:collinear}, the absolute minima $P_A =
0$ never appear as isolated points in the oscillograms, but always
form continuous lines (valleys of zero probability).  Such a property
(degeneracy of minima) is lifted for non-zero $\Dmq_{21}$~\cite{AMS2}.
This is unlike for the absolute maxima, such as the MSW mantle peak or
the parametric resonance peak in the core region, where instead the
value $P_A = 1$ is reached only at a few isolated points. This feature
is a consequence of the symmetry of the matter density profile of the
Earth, and can be readily understood in the following way.

The condition for the absolute minimum, $P_A = 0$, or $S_{12} = 0$,
can be written as
\begin{equation} \label{eq:abmin}
    \Re S_{12}(L, 0) = 0\,,\qquad \Im S_{12}(L, 0) = 0\,,
\end{equation}
and for a generic profile the absolute minima are found as the points
where the curves corresponding to the two conditions
in~\eqref{eq:abmin} intersect. However, due to the symmetry of the
Earth's matter density profile, the condition $\Re S_{12}(L) = 0$ is
satisfied automatically for all values of $E_\nu$ and $\Theta_\nu$.
Therefore the zeros of $P_A$ simply coincide with the contour curves
$\Im S_{12}(L) = 0$.

The absolute maxima, $P_A \equiv |S_{12}|^2 = 1$, are realized when
$|S_{11}|^2 = 0$, or
\begin{equation} \label{eq:abmax}
    \Re S_{11}(L, 0) = 0\,, \qquad \Im S_{11}(L, 0) = 0\,.
\end{equation}
Since in general $\Re S_{11}$ and $\Im S_{11}$ are independent and
non-trivial functions of $E_\nu$ and $\Theta_\nu$, the contours that
correspond to equalities in~\eqref{eq:abmax} do not coincide. The
absolute maxima occur only at the intersections of the these contours,
which explains why such maxima are isolated points.

Another interesting feature of the transition probability $P_A$ is
that its dependence on the distance $x$ along a given trajectory
exhibits peculiar symmetry properties for trajectories, corresponding
to the absolute minima and maxima of $P_A(L)$. Specifically,
\begin{myitemize}
  \item For trajectories, corresponding to the absolute minima
    ($P_A(L) = 0$),
    \begin{equation}
	P_A(L-x) = P_A(x)\,.
    \end{equation}
    That is, $P_A$ is symmetric with respect to the midpoint of the
    trajectory $x = L/2$: $P_A(L/2 + z) = P_A(L/2 - z)$, $z\le L/2$.
    
  \item For trajectories, corresponding to the absolute maxima
    ($P_A(L) = 1$),
    \begin{equation}
	P_A(L-x) = 1 - P_A(x)\,.
    \end{equation}
    This implies that in the middle of the trajectory $P_A(L/2) = 1/2$
    and the function $P'\equiv P_A - 1/2$ is antisymmetric with
    respect to the midpoint: $P'(L/2 + z) = - P'(L/2 - z)$.
\end{myitemize}
The proof is straightforward. Due to the symmetry of the density
profile with respect to the midpoint of the neutrino trajectory, we
have for any point $x$ on the trajectory
\begin{equation} \label{eq:tinv}
    S(L-x, 0) = S^T(L, x),
\end{equation}
which is essentially a consequence of T-invariance of 2-flavour
neutrino oscillations~\cite{Tviol}. Then, from the definition of the
evolution matrix one finds $S(L, 0) = S(L,x) \, S(x, 0)$, which can be
rewritten (using the unitarity of $S$) as $S(L, x) = S(L, 0)\, S(x,
0)^{\dagger}$.  Plugging the latter relation into~\eqref{eq:tinv}, we
obtain
\begin{equation} \label{eq:relss}
    S(L-x,0) = S^*(x, 0) \, S(L, 0)\,.
\end{equation}

For the absolute minima, $P_A=0$, the evolution matrix $S(L, 0)$
should be diagonal, and therefore we obtain from Eq.~\eqref{eq:relss}
\begin{equation}
    S(L-x,0) = S^*(x, 0)
    \begin{pmatrix}
	e^{i\phi_L} & 0 \\
	0 & e^{- i\phi_L}
    \end{pmatrix}.
\end{equation}
Consequently,
\begin{equation}
    P_A(L-x) = \big| S_{12}^*(x, 0) e^{-i\phi_L} \big|^2 = P_A(x)\,.
\end{equation}

For the absolute maxima, $P_A(L) = 1$, the matrix $S(L, 0)$ should be
off-diagonal (see Eq.~\eqref{eq:abmax}) with pure imaginary elements
due to the symmetry of the matter density profile. Therefore, 
using~\eqref{eq:relss}, one finds
\begin{equation}
    S(L-x, 0) = S^*(x, 0)
    \begin{pmatrix}
	0 & -i \\
	-i & 0
    \end{pmatrix}\,,
\end{equation}
and consequently,
\begin{equation}
    P_A(L-x) = \big|S_{11}^*(x, 0) (-i) \big|^2 = 1 - P_A(x) \,.
\end{equation}

Notice that, according to Fig.~\ref{fig:collinear}, there are no local
minima with $P_A \ne 0$.


\subsection{Interpretation of the oscillation pattern for core
  crossing trajectories}
\label{sec:graphical}

The analysis presented in the previous subsections allows one to give
a complete physics interpretation of neutrino oscillations in the
Earth. Essentially the whole oscillatory pattern that includes the
ridges, absolute and local maxima, saddle points and zeros, can be
understood on the basis of different realizations of the amplitude and
phase conditions.
We illustrate neutrino oscillations inside the Earth for core crossing
trajectories by Figs.~\ref{fig:evol-050} and~\ref{fig:evol-125}, and
the corresponding graphical representation of the oscillations is
given in Figs.~\ref{fig:graph}b--\ref{fig:graph}f. For comparison, in 
Fig.~\ref{fig:graph}a we also show the graphical representation of the
oscillations for a trajectory crossing only the mantle of the Earth.

\PAGEFIGURE{
  \includegraphics[width=145mm]{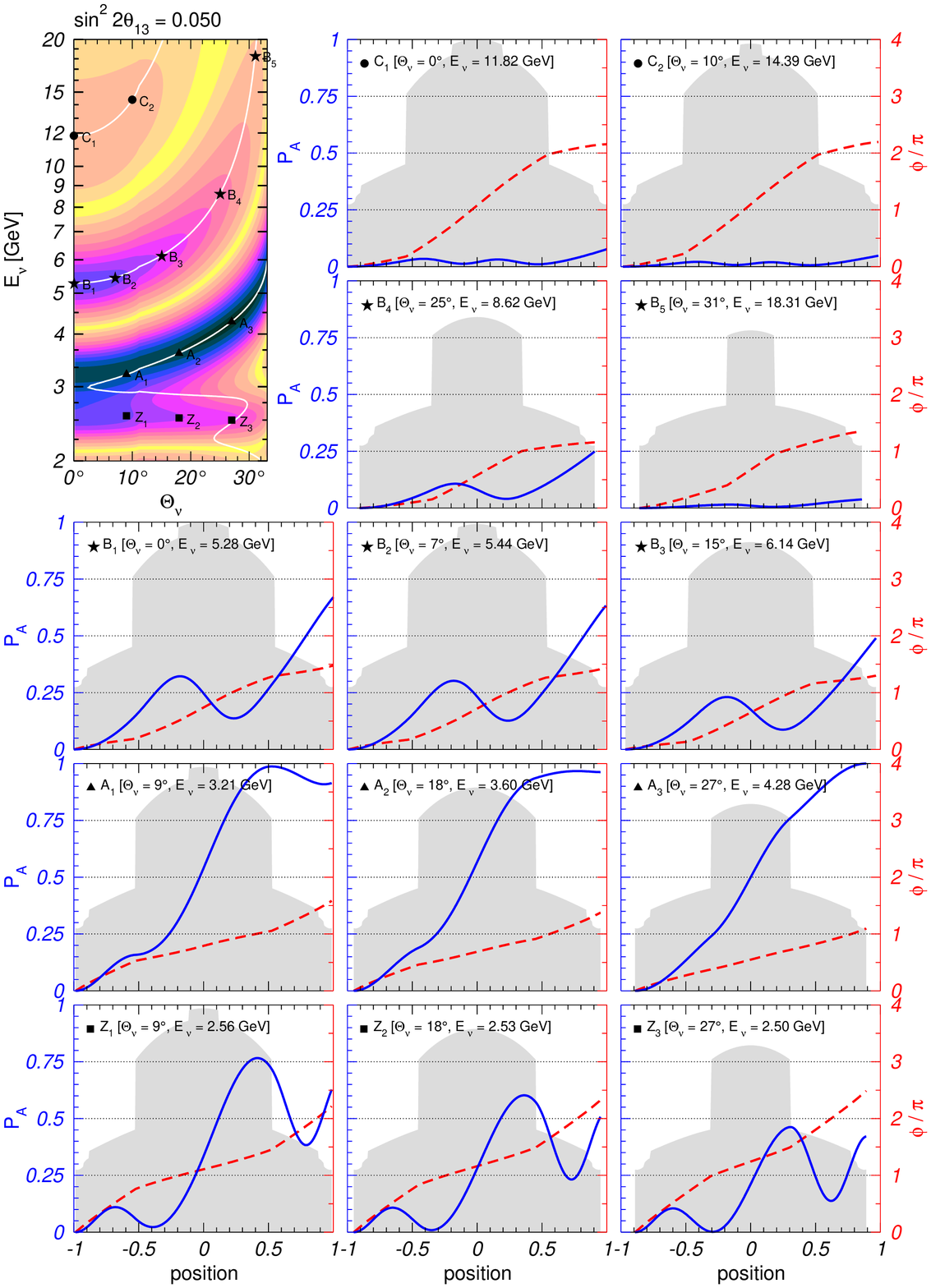}
  \caption{\label{fig:evol-050}%
    Dependence of the probability $P_A$ (solid blue), half-phase
    $\phi$ (dashed red) and the Earth density (gray shade) on the
    position along the neutrino trajectory, for $\sin^2 2\theta_{13} =
    0.05$ and different values of $\Theta_\nu$ and $E_\nu$. The
    corresponding points of the oscillogram, together with the curves
    of the amplitude (collinearity) condition, are shown in the
    upper-left panel.} 
  }

\PAGEFIGURE{
  \includegraphics[width=145mm]{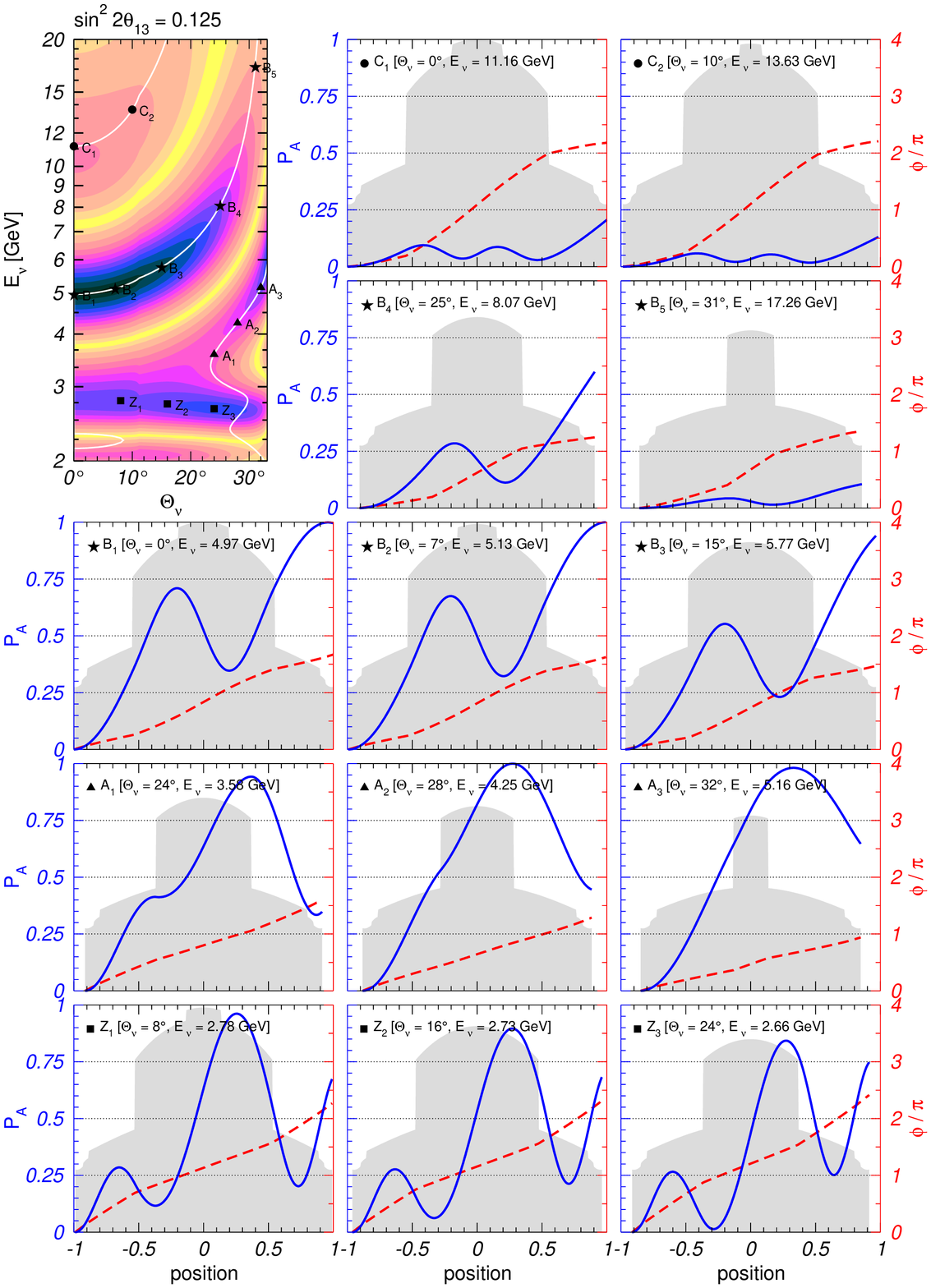}
  \caption{\label{fig:evol-125}%
    Same as in Fig.~\ref{fig:evol-050}, but for $\sin^2 2\theta_{13} =
    0.125$.}
  }

\textitem{Core ridge.} The core resonance ridge is located at $E_\nu
\sim (2.5 - 2.8)$ GeV.  It is of the MSW resonance nature, but is
situated below the MSW resonance line in the core: $E_\nu <
E_{R}(\Theta_\nu)$, the reason being that the values of the
oscillation half-phase are different from $\pi/2$. The ridge does not
coincides with any curve corresponding to the phase or amplitude
condition.  At one point on the ridge there is an intersection of the
curves $\Im\beta_1 = 0$ (the mantle half-phase $\phi_1 = \pi$) and the
collinearity condition, as well as of the $\Re\alpha_2 = 0$ curve,
corresponding to the core half-phase $\phi_2 = \pi/2$ (see
Fig.~\ref{fig:collinear}). This crossing point corresponds to the
local maximum with zero mantle effect and maximal oscillation
amplitude in the core.

As follows from Figs.~\ref{fig:evol-050} and~\ref{fig:evol-125}, the
main contribution to the oscillation probability comes from the 
MSW-enhanced oscillations in the core, though the mantle contribution
is not negligible. The detailed picture depends on the value of
$\sin^2 2\theta_{13}$.

For the inner core trajectories with $\Theta_\nu \sim 0$, the core
phase $\phi_2 > \pi/2$, and it decreases along the ridge with
$\Theta_\nu$ increasing.  For small mixing angles, \eg\ $\sin^2
2\theta_{13} \leq 0.05$ (see Fig.~\ref{fig:evol-050}, lower panels),
the contributions of the two mantle layers to $P_A$ practically cancel
each other, so that the probability is determined by the oscillations
in the core.
The oscillation phase in the core, $\phi_2$, weakly decreases with the
increase of $\Theta_\nu$, in spite of a substantial decrease of the 
trajectory length. The phase acquired in the core can be written as
\begin{equation} \label{eq:core-ridge}
    \phi_2 = \frac{\pi L_2 (\Theta_\nu)}{l_m (\bar{V}(\Theta_\nu))}
    \approx \text{const} \,,
\end{equation}
where $\bar{V}(\Theta_\nu)$ is the value of the potential averaged 
along a given trajectory and $L_2 (\Theta_\nu) = 2R_c \cos \Theta_\nu$
is the neutrino path length inside the core.  The ridge is below the
resonance and therefore the oscillation length in matter
\emph{decreases} sharply (for small mixing) with the potential and 
also with energy.  With the increase of $\Theta_\nu$ the length of the
core-trajectory $L_2$ decreases, but simultaneously, the oscillation
length decreases since the average potential decreases according to
the PREM profile. These two dependencies compensate each other in
Eq.~\eqref{eq:core-ridge} leading to weak change of $\phi_2$.

At $\Theta_\nu = 27^\circ$ (point $Z_3$) the phase reaches $\phi_2 =
\pi/2$ and furthermore $\phi_1 = \pi$. This corresponds to pure core
effect. Notice that with increase of $\Theta_\nu$ the average density
decreases and the depth of oscillations becomes smaller. In
Fig.~\ref{fig:graph}b we show the graphical representation of
evolution with parameters from core-ridge (point $Z_3$).  We show the
precession cones in the mantle and in the core in the points close to
border between the mantle and core. Shift of the evolution trajectory
from the cone surfaces is due to density change and violation of
adiabaticity.

For $\sin^2 2\theta_{13} = 0.05$ another intersection of lines of the
amplitude condition and the phase conditions $\Re \alpha_2 = 0$ (core
half-phase $\pi/2$) occurs at $E_\nu = 3.3$ GeV and $\Theta_\nu =
10^\circ$. Evolution for this configuration is shown in
Fig.~\ref{fig:evol-050} point $A_1$. Notice that this point is in the
ridge $A$.  Here transitions in both mantle layers are non-zero but
have opposite sign and cancel each other. So, the whole effect due to
MSW resonance enhancement in core.

For large 1-3 mixing, \eg\ $\sin^2 2\theta_{13} = 0.125$, there is
substantial interplay of the core and mantle oscillation effects (see
Fig.~\ref{fig:evol-125}). For $\Theta_\nu \sim 0$, we find $\phi_1
\sim \phi_2 < \pi$; contributions from two mantle layers interfere
constructively and the total mantle contribution is comparable with
the core contribution (the later is resonantly enhanced).  With
increase of $\Theta_\nu$, the phase $\phi_1$ increases reaching $\pi$,
whereas $\phi_2$ decreases. Mantle contribution decreases, and
moreover, effects of two mantle layers start to compensate each other.
At $\Theta_\nu = 29^\circ$ the condition of local maximum are
satisfied.

Notice that for large $\sin^2 2\theta_{13}$ the decrease of $l_m$ with
$\bar{V}$ is weaker and therefore it can not compensate decrease of
the trajectory length. The phase stays constant if also the energy
decreases. Consequently, for $\sin^2 2\theta_{13} > 0.05$ the energy
of ridge line decreases with increase of $\Theta_\nu$. Even stronger
dependence of the energy on $\Theta_\nu$ can be seen in the case of
constant density in the core.

Recall that in this domain of low energies the oscillations due to 1-2
splitting become important~\cite{AMS2}.

\PAGEFIGURE{
  \includegraphics[width=115mm]{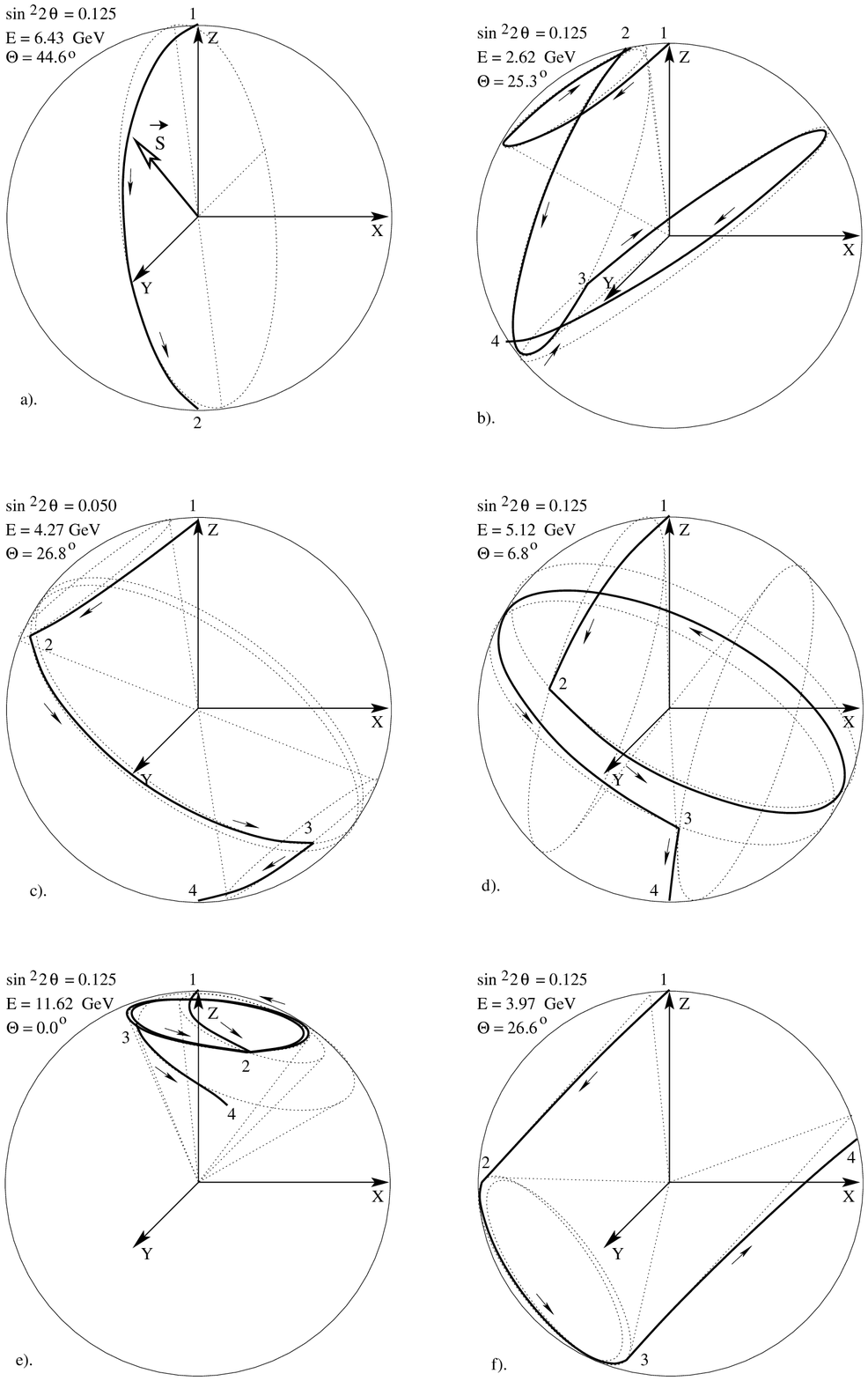}
  \caption{\label{fig:graph}%
    Graphical representation of neutrino evolution in the Earth. Shown
    are the trajectories of the neutrino vector $\vec{s}$ in the
    $(X,\, Y,\, Z)$ space (solid lines), for different values of
    $\theta_{13}$ and different points in the oscillograms.  The
    different panels correspond to a) the MSW peak in mantle; b) the
    point $Z_2$ in Fig.~\ref{fig:evol-125}; c) the point $A_3$ in
    Fig.~\ref{fig:evol-050}; d) the point $B_2$ in
    Fig.~\ref{fig:evol-125}; e) the point $C_1$ in
    Fig.~\ref{fig:evol-125}; f) near saddle point $A_1$ in
    Fig.~\ref{fig:evol-125}. We also show the precession cones (dotted
    lines) in the mantle and core points close to the border between
    the mantle and core. Sections 1-2, 2-3 and 3-4 indicate evolution
    in the first mantle layer, the core and the second mantle layer
    correspondingly.}
  }

\textitem{}The parametric ridges differ by the oscillation phase
acquired in the core, $\phi_2$.

\textitem{Ridge A.} The phase in the core $\phi_2 \lesssim \pi$. This
ridge lies in between the core resonance (at $\Theta_\nu \sim
0^\circ$) and the mantle resonance regions. At the border between the
core and the mantle domains of the oscillogram, $\Theta_\nu =
33.1^\circ$, this ridge merges with the MSW resonance peak in the
mantle.

This ridge is large (has the largest area of probability close to 1)
for small 1-3 mixing: $\sin^2 2\theta_{13} \lesssim 0.05$.  As we
mentioned above, for $\Theta_\nu < 10^{\circ}$ and $E_\nu \sim 3$ GeV
the phase in the core $\phi_2 < \pi/2$, the mantle effect is small and
two mantle layer contributions cancel each other.  So, here we deal
with resonance enhancement of oscillations in core.  With increase of
$E_\nu$ and $\Theta_\nu$ along the ridge both phases $\phi_1$ and
$\phi_2$ decrease. The core and two mantle contributions add
constructively leading to large probability.  For example, for the
point $A_3$ (Fig~\ref{fig:evol-050}) $\pi_1 \approx \pi_2 \approx
\pi/3$ and maximal probability build up by three comparable
contributions. The graphical representation of this evolution is shown
Fig.~\ref{fig:graph}c.
For the core part we show in some panels two precession cones,
corresponding to the beginning and the end of the core section. The
difference of these two cones reflects the effects of the non-constant
density distribution.

For large 1-3 mixing $\sin^2 2\theta_{13} \geq 0.075$, the region of
maximal transition shifts to the mantle domain. Furthermore, the
saddle point appear between the MSW resonance and the parametric
resonance regions. In the saddle point $A_1$ (Fig.~\ref{fig:evol-125})
the phases in the core and the mantle $\phi_1 \approx \phi_2 \approx
\pi/2$, correspond to maximal oscillatory factors. However the
contributions from the two mantle layers have opposite sign and cancel
each other. The corresponding graphical representation is shown in
Fig.~\ref{fig:graph}f.

With increase of $\Theta_\nu$ the phases $\phi_1$ and $\phi_2$
decrease and cancellation becomes weaker.

In the ridge $A$ for small 1-3 mixing the lines of the phase and
amplitude conditions almost coincide that ensures stability of
enhancement in large area. This stability can be understood also in
the following way. With increasing $\Theta_\nu$, the core segment of
the trajectory becomes shorter and the mantle ones become longer. This
is compensated by an increase of energy along the ridge. Indeed, since
the ridge $A$ is between the core and the mantle MSW resonance
energies, with an increase of energy the oscillation length in the
core (above the MSW resonance) decreases, whereas the oscillation
length in the mantle (below the mantle MSW resonance) increases, so
that the oscillation phases change rather weakly. Along the ridge, in
the direction of larger $\Theta_\nu$ the depth of oscillations in the
core decreases, and this is compensated by an increase of the depth in
the mantle. All this produces a long ridge and ensures its stability.

\textitem{Ridge B.} This ridge is situated at $E_\nu \geq 5$ GeV. For
the smallest energies in the ridge and $\Theta_\nu \sim 0$ the
half-phase in the core $\phi_2 \sim (1.2 - 1.3) \pi$, so that
oscillations in the core give substantial contribution effect (see
Figs.~\ref{fig:evol-050} and~\ref{fig:evol-125}).  In the range $E_\nu
= (5 - 6)$ GeV the parametric enhancement of oscillations with
significant interplay of the core and mantle oscillation effects is
realized.  In Fig.~\ref{fig:graph}d we show graphical representation
for the point $B_2$ (Fig.~\ref{fig:evol-125}).

With increase of $\Theta_\nu$ and $E_\nu$ the phase $\phi_2$ decreases
due to decrease of the trajectory length $L_2$ (notice that here $l_m$
decreases too approaching $l_0$ but much weaker than $L_2$.) In
contrast, the phase in the mantle increases. For $E_\nu = (6 - 7)$ GeV
one finds $\phi_2 \approx \pi$ an the core effect is absent, so that
transition occurs due to the oscillations in mantle only.

For $E_\nu > 20$ GeV the equality $\phi_2 \sim \phi_1 \sim \pi/2$ is
realized. Indeed, the lines $\Re \alpha_1$ and $\Re \alpha_2$ as well
as lines of the amplitude and phase conditions cross in the point
$E_\nu \approx 21$ GeV and $\Theta_\nu = 31^\circ$. Here maximal
enhancement of the transition probability for given values of mixing
angles occurs~\cite{AMS1}. It is described by Eq.~\eqref{eq:saddle}.

With an increase of both $E_\nu$ and $\Theta_\nu$ along the ridge the
oscillation amplitudes decrease both in the core and in the mantle
layers. The core-oscillation depth is suppressed stronger and
therefore the region of relatively high probability lies in the same
energy domain as the region of high probability in the mantle.

In this ridge a qualitative picture of oscillations does not depend on
$\sin^2 2\theta_{13}$.

\textitem{Ridge C.} The ridge is located at $E_\nu > 11$ GeV in the
matter dominated region where mixing and consequently oscillation
depth are suppressed. For $\Theta_\nu \sim 0$ the half-phase in the
core equals $\phi_2 \sim 1.8 \pi$. Here main contribution to
probability is due to oscillations in mantle whereas the core gives
small (negative) contribution. We show graphical representation of
evolution that corresponds to the point $C_1$ in
Fig.~\ref{fig:graph}e.

With increase of $\Theta_\nu$ and $E_\nu$ along the ridge the phase
$\phi_2$ decreases and the $E_\nu > 100$ GeV reaches $3 \pi/2$,
whereas the mantle phase is about $\phi_1 \sim \pi/2$. Here, as in the
case of ridge $B$, the parametric enhancement of oscillations is
realized~\cite{AMS1} as described by the Eq.~\eqref{eq:saddle}.

\textitem{}Notice that qualitative features of the high energy part of
ridge $B$ and ridge $C$ do not depend on $\theta_{13}$ since both are
in the $\theta_{13}$ factorization region where $P_A \propto \sin^2
2\theta_{13}$ (see Sec.~\ref{sec:13dep}).


\section{Approximate analytic description of neutrino oscillations in
  the Earth}
\label{sec:approx}

In this section we develop an approximate analytic approach to
2-flavor neutrino oscillations in the Earth. To this end, we employ a
perturbation theory in the deviation of the density profile from that
represented by layers of constant densities. 
This approach has been first developed in Ref.~\cite{Lisi:1997yc} for
the description of the solar neutrino oscillations in the earth. Here
we apply it to the study of atmospheric neutrinos.
The approximation turns out to work extremely well in spite of the
fact that the variations of the neutrino potential inside the Earth
layers can be large $|\Delta V|/ \bar{V} \sim 0.3$. The high accuracy
of our approach is related to the symmetry of the density profile of
the Earth.


\subsection{Perturbation theory in $\Delta V$}
\label{sec:formalism}

Let us first consider the case of one layer of relatively weakly
varying density and represent the matter-induced potential of
neutrinos $V(x)$ along a given trajectory as the sum of a constant
term $\bar{V}$ and a perturbation $\Delta V(x)$:
\begin{equation}
    V(x) = \bar{V} + \Delta V(x) \, .
\end{equation}
Correspondingly, the Hamiltonian of the system can be written as the
sum of two terms:
\begin{equation}
    H(x) = \bar{H} + \Delta H(x)\,,
\end{equation}
where
\begin{equation}
    \bar{H} \equiv \bar\omega
    \begin{pmatrix}
	-\cos 2\bar\theta & \sin 2\bar\theta \\
	\hphantom{-}\sin 2\bar\theta & \cos 2\bar\theta
    \end{pmatrix},
    \qquad
    \Delta H \equiv \frac{\Delta V(x)}{2}
    \begin{pmatrix}
	1 & \hphantom{-}0 \\
	0 & -1
    \end{pmatrix}\,.
\end{equation}
Here $\bar\theta = \theta_m(\bar V)$ is the mixing angle in matter and
$\bar\omega = \omega(\bar V)$ is half of the energy splitting
(half-frequency) in matter, both with the average potential $\bar V$.
Throughout this chapter we will denote by $\bar{S}(x)$ the evolution
matrix of the system for the constant density case $H(x) = \bar{H}$.
The explicit expression for $\bar{S}(x)$ is given by
Eq.~\eqref{eq:Ubar} with $\theta_m = \bar\theta$.

For matter of varying density, we seek the solution of the evolution
equation~\eqref{eq:smatreq} in the form
\begin{equation} \label{eq:exp2}
    S(x) = \bar{S}(x) + \Delta S(x), \qquad
    \Delta S(x) = -i \, \bar{S}(x) \, K_1(x)\,,
\end{equation}
where $K_1(x)$ satisfies $|K_1(x)_{ab}| \ll 1$. Inserting
Eq.~\eqref{eq:exp2} into Eq.~\eqref{eq:smatreq}, we find the following
equation for $K_1(x)$ to the first order in $\Delta H(x)$ and
$K_1(x)$:
\begin{multline} \label{eq:diff}
    \frac{dK_1(x)}{dx} =
    \bar{S}^\dagger(x) \, \Delta H(x) \, \bar{S}(x) =
    \frac{\Delta V(x)}{2} \LT\lbrace
    -\cos2\bar\theta
    \begin{pmatrix}
	-\cos2\bar\theta & \sin2\bar\theta \\
	\hphantom{-}\sin2\bar\theta & \cos2\bar\theta
    \end{pmatrix}
    \RT.
    \\[2mm]
    \LT.
    + \sin2\bar\theta
    \begin{pmatrix}
	\sin2\bar\theta & \hphantom{-}\cos2\bar\theta \\
	\cos2\bar\theta & -\sin2\bar\theta
    \end{pmatrix} \cos2\phi(x)
    + \sin2\bar\theta
    \begin{pmatrix}
	0 & -i \\
	i & \hphantom{-}0
    \end{pmatrix} \sin2\phi(x)
    \RT\rbrace \,.
\end{multline}
The first term does not contribute to $S \equiv S(L)$ since $\langle
\Delta V \rangle \equiv \int \Delta V(x) dx = 0$, and
Eq.~\eqref{eq:diff} can be immediately integrated:
\begin{multline}
    K_1(L) = \frac{1}{2} \sin2\bar\theta \LT\lbrace
    \begin{pmatrix}
	\sin2\bar\theta & \hphantom{-}\cos2\bar\theta \\
	\cos2\bar\theta & -\sin2\bar\theta
    \end{pmatrix}
    \int_0^L \Delta V(x) \, \cos2\phi(x) \, dx
    \RT.
    \\
    \LT. +
    \begin{pmatrix}
	0 & -i \\
	i & \hphantom{-}0
    \end{pmatrix}
    \int_0^L \Delta V(x) \, \sin2\phi(x) \, dx
    \RT\rbrace \,.
    \label{eq:K1}
\end{multline}
It is convenient to introduce the new variable $z = x - L/2$, which
measures the distance from the midpoint of the neutrino trajectory.
Then from~\eqref{eq:K1} we obtain
\begin{equation}
    \label{eq:DeltaU}
    \Delta S \equiv \Delta S(L) = -i \, \sin2\bar\theta
    \LT\lbrace
    \begin{pmatrix}
	\sin2\bar\theta & \hphantom{-}\cos2\bar\theta \\
	\cos2\bar\theta & -\sin2\bar\theta
    \end{pmatrix}
    \Delta I +
    \begin{pmatrix}
	0 & -i \\
	i & \hphantom{-}0
    \end{pmatrix}
    \Delta J
    \RT\rbrace \,,
\end{equation}
where
\begin{gather}
    \label{eq:DeltaVcs}
    \Delta I \equiv \frac{1}{2}
    \int_{-L/2}^{L/2} \Delta V(z) \, \cos(2\bar\omega z) \, dz \,,
    \qquad
    \Delta J \equiv \frac{1}{2}
    \int_{-L/2}^{L/2} \Delta V(z) \, \sin(2\bar\omega z) \, dz \,.
\end{gather}
In these integrals, $\Delta V(z) \equiv \Delta V(x(z))$ and $x(z) = z
- L/2$. Obviously, $\Delta J$ vanishes if the perturbation $\Delta
V(z)$ is symmetric with respect to the midpoint of the trajectory.
Analogously, $\Delta I$ vanishes if $\Delta V(z)$ is antisymmetric.
The expression for $S$ defined in Eq.~\eqref{eq:exp2} with $\Delta S$
given in Eqs.~\eqref{eq:DeltaU} and~\eqref{eq:DeltaVcs} is equivalent
to Eqs.~(13--16) obtained in Ref.~\cite{Lisi:1997yc} in the context of
solar neutrino oscillations.

Let us now consider the issue of the unitarity of the obtained
evolution matrix that can be important for numerical calculations.
Since $\bar{S} \equiv \bar{S}(L)$ is unitary, we have
\begin{equation} \label{eq:SSunitar}
    S^{\dagger} S = I +
    \Delta S^{\dagger} \bar{S} + \bar{S}^{\dagger} \Delta S
    + \Delta S^{\dagger} \Delta S.
\end{equation}
The second and third terms on the RHS of this equality cancel each
other thus ensuring the unitarity of $S$ at the first order in $\Delta
S$. In order to prove this it is convenient to parametrize the
evolution matrix $\bar{S} $ and the perturbation $\Delta
S$~\eqref{eq:DeltaU} as in Eq.~\eqref{eq:U2}:
\begin{align}
    \label{eq:Ucomp}
    Y &= \cos\phi \,,
    & \mathbf{X} &= \sin\phi
    \LT( \sin2\bar\theta ,\; 0 ,\; -\cos2\bar\theta \RT) \,,
    \\[2mm]
    \label{eq:DUcomp}
    \Delta Y &= 0,
    & \Delta \mathbf{X} &= \varepsilon \LT(
    \cos2\bar\theta \, \cos\xi ,\; \sin\xi ,\;
    \sin2\bar\theta \, \cos\xi \RT) \,,
\end{align}
where $\phi \equiv \phi(L)$ and we have introduced
\begin{equation}
    \varepsilon = \sin2\bar\theta \, \sqrt{\Delta I^2 + \Delta J^2} \,,
    \qquad
    \xi = \arg( \Delta I + i\, \Delta J ) \,.
\end{equation}
It is easy to see that $\Delta S^\dagger \bar{S} + \bar{S}^\dagger
\Delta S = Y \, \Delta Y + \mathbf{X} \cdot \Delta\mathbf{X} = 0$. The
last term in Eq.~\eqref{eq:SSunitar}, $\Delta S^{\dagger} \Delta S$,
violates the unitarity condition, however it is of order
$\varepsilon^2$ and therefore it does not break consistency of our
approximation. However, for practical purposes it would be useful to
have an expression for $S$ which is \emph{exactly} unitary regardless
of the size of the perturbation, even if for large perturbations the
formalism is no longer accurate. In order to do this, we first rewrite
Eq.~\eqref{eq:DeltaU} as follows:
\begin{equation}
    \Delta S = \varepsilon \, S' \,,
    \qquad
    S' = -i
    \LT\lbrace
    \begin{pmatrix}
	\sin2\bar\theta & \hphantom{-}\cos2\bar\theta \\
	\cos2\bar\theta & -\sin2\bar\theta
    \end{pmatrix}
    \cos\xi +
    \begin{pmatrix}
	0 & -i \\
	i & \hphantom{-}0
    \end{pmatrix}
    \sin\xi
    \RT\rbrace \,,
\end{equation}
and then we perform the following replacement in the expression for
$S$:
\begin{equation} \label{eq:Uimproved}
    S = \bar S + \varepsilon \, S'
    \quad\longrightarrow\quad
    S = \cos\varepsilon \, \bar S + \sin\varepsilon \, S' \,.
\end{equation}
Note that both $S'$ and $\bar S$ are unitary matrices, and that due to
their specific form the combination on the right-hand-side of
Eq.~\eqref{eq:Uimproved} is exactly unitary. In our computations we
use the exactly unitary matrix~\eqref{eq:Uimproved}.


\subsection{Application: mantle-only crossing trajectories}
\label{sec:mantle}

We shall now use the formalism of the previous subsection to derive an
explicit formula for the transition probability for mantle-only
crossing neutrino trajectories. Since the density profile is symmetric
with respect to the midpoint of the trajectory, the term $\Delta J$ is
absent. From Eqs.~\eqref{eq:Ubar}, \eqref{eq:DeltaU}
and~\eqref{eq:Uimproved} we immediately get
\begin{equation} \label{eq:probab}
    P_A = \LT[ \cos\varepsilon \, \sin2\bar\theta \, \sin\phi
    + \sin\varepsilon \, \cos2\bar\theta \RT]^2
    \approx \sin^22\bar\theta \, \LT[ \sin\phi + \Delta I \,
    \cos2\bar\theta \RT]^2,
\end{equation}
where $\varepsilon \equiv \sin2\bar\theta \, \Delta I$ and $\phi
\equiv \phi(L) = \bar\omega L$. Here the first term in the square
brackets describes oscillations in constant density matter with
average potential $\bar{V}_1$.
In order to obtain an explicit formula for $\Delta I$, we approximate
the matter density profile along the neutrino trajectory by a
parabola:
\begin{equation} \label{eq:parab}
    \Delta V(z) \approx
    V''_1 \LT[ \LT( \frac{z}{L} \RT)^2 - \frac{1}{12} \RT].
\end{equation}
The average value $\bar{V}_1$ and the coefficient $V''_1$ depend only
on the nadir angle $\Theta_\nu$, and are shown in
Fig.~\ref{fig:profile}. Inserting the expression for $\Delta V(z)$
into Eq.~\eqref{eq:DeltaVcs} and integrating by parts, we obtain
\begin{equation} \label{eq:phi-int}
    \Delta I = \frac{V''_1 L}{12} f(\phi), \qquad
    f(\phi) \equiv \frac{3\phi \cos\phi
      + (\phi^2 - 3) \sin\phi}{\phi^3} \,.
\end{equation}
The function $f(\phi)$ has the following features: for $\phi
\rightarrow 0$ (outer trajectories), one has $f(\phi)\rightarrow
-\phi^2/15$; for $\phi=\pi/2$ which corresponds to the maximum
transition probability for a given mixing angle in matter, $f(\phi)
\approx - 0.13$; the function $|f(\phi)|$ reaches its maximum,
$f(\phi) \approx - 0.31$, at $\phi \simeq 1.07\pi$.
The function $f(\phi)$ changes its sign at $\phi \sim 1.8\pi$, and for
large $\phi$ it behaves as $f(\phi) \sim \sin\phi / \phi$.

From Eqs.~\eqref{eq:probab} and~\eqref{eq:phi-int} it follows that:
\begin{myitemize}
  \item in the zeroth approximation, the transition probability is
    given by the standard oscillation formula for matter of constant
    density, with the oscillation amplitude determined by the mixing
    angle $\bar{\theta} = \theta_m(\bar{V}_1)$ and the phase $\phi =
    \omega(\bar{V}_1)L$;
    
  \item the lowest-order correction to $P_A$ vanishes at the MSW
    resonance, \ie\ along the curve $E_R(\Theta_\nu)$. The correction
    changes its sign at the resonance, being positive below it
    ($E_\nu<E_R(\Theta_\nu))$ and negative above it;
    
  \item the largest corrections correspond to the trajectories with
    $\Theta_\nu \sim 37^\circ$ (\ie\ passing close to the core) and
    the phase $\phi \sim \pi$. Indeed, $f(\phi)$, $|V''_1|$ and $L$
    are all maximal for the deepest mantle trajectories
\end{myitemize}

The transition probability defined by Eqs.~\eqref{eq:probab}
and~\eqref{eq:phi-int} can be cast into the form resembling formally
the standard oscillation probability in vacuum or in matter of
constant density:
\begin{equation}
    P_A = \sin^2 2\theta' \, \sin^2 \phi'\,,
\end{equation}
where
\begin{equation} \label{eq:notat1}
    \sin^2 2\theta' \equiv (a^2 + b^2) \sin^2 2 \bar\theta, \qquad
    \phi' = \phi + \arctan(a/b)
\end{equation}
with
\begin{equation} \label{eq:notat2}
    a = \frac{V''_1 L}{4 \phi^2} \cos 2\bar{\theta}\,, \qquad
    b = 1 + V''_1 L \frac{\phi^2 - 3}{12 \phi^3}\cos 2\bar{\theta}\,.
\end{equation}
This representation has the advantage that it factorizes the
transition probability into a smooth function of neutrino energy and
nadir angle, $\sin^2 2\theta'$, and the oscillating factor $\sin^2
\phi'$ lying between 0 and 1. In particular, zero probability curves
in the oscillograms are given by the condition $\phi'(\Theta_\nu,
E_\nu) = \pi k$.
Factorization shows immediately continuous character of lines of zero
probability. Note that the effective oscillation amplitude $\sin^2
2\theta'$ is not actually a sine squared of any angle and should be
simply understood as a notation defined in Eq.~\eqref{eq:notat1}; in
particular, it is not bounded from above by unity.

\FIGURE[!t]{
  \includegraphics[width=145mm]{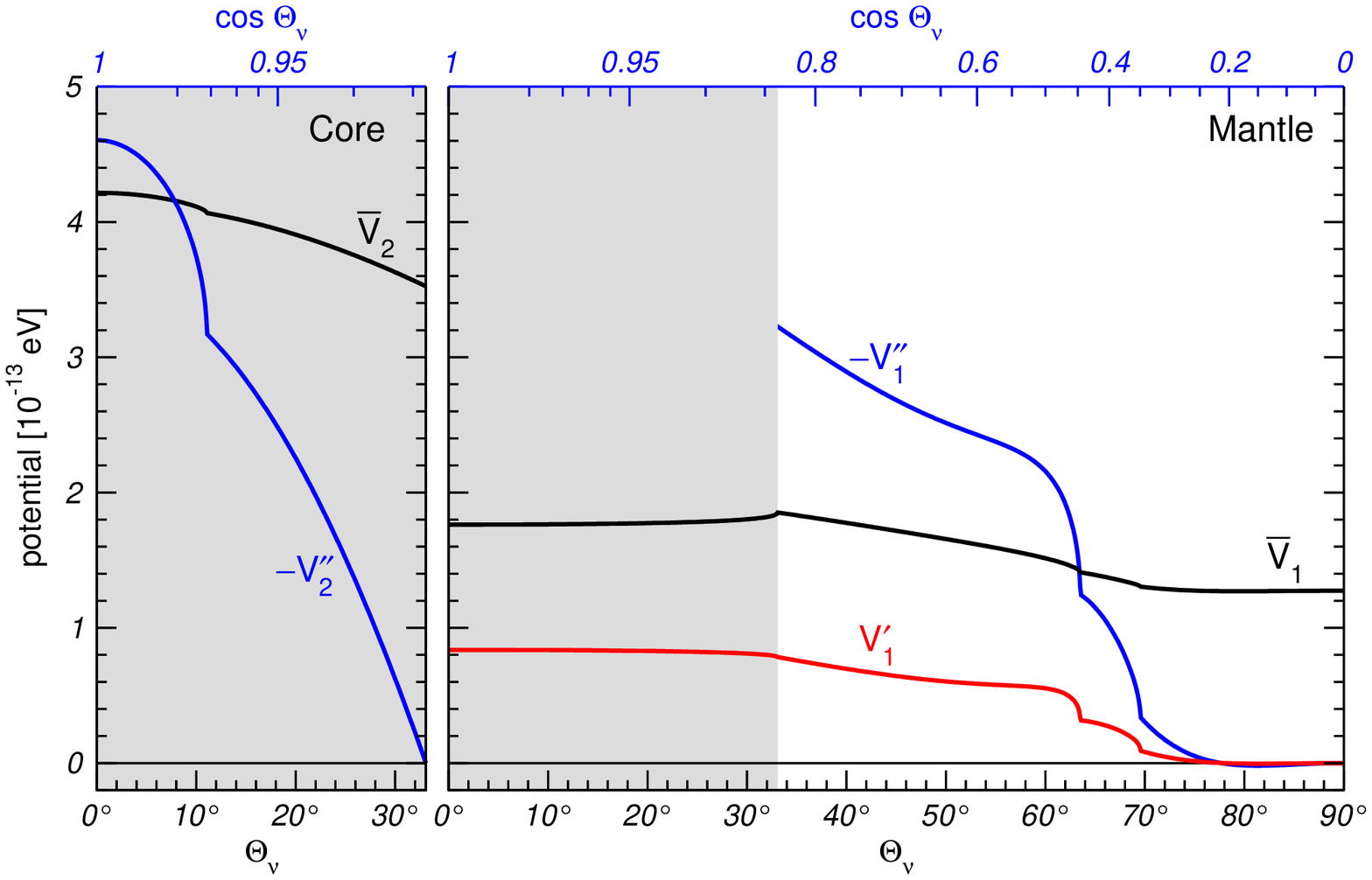}
  \caption{ \label{fig:profile} %
    Dependence of the average potentials ($\bar{V}_1$, $\bar{V}_2)$
    (black) as well as the coefficients $V'_1$ (red) and ($-V''_1$,
    $-V''_2$) (blue) on the nadir angle $\Theta_\nu$.}
  }

\FIGURE[!t]{
  \includegraphics[width=145mm]{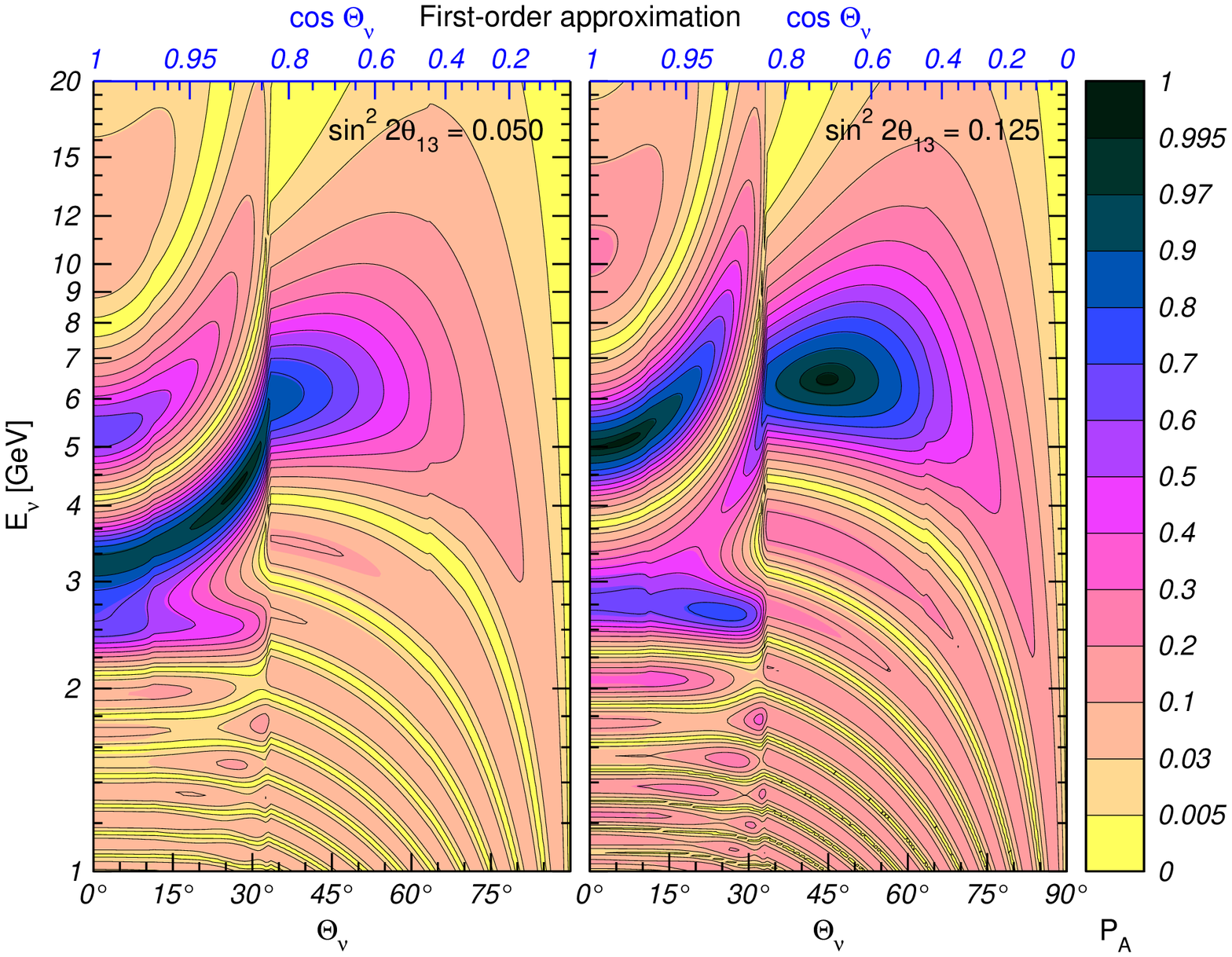}
  \caption{\label{fig:apx-impro}%
    Contour plot of the probability $P_A$ for the PREM density profile
    (colored regions; grayscale on black-and-white printouts) and for
    our analytic approximation including first-order corrections
    (black curves).}
  }


\subsection{Application: core-crossing trajectories}
\label{sec:core}

For trajectories crossing the Earth's core, the evolution matrix can
be factorized as
\begin{equation} \label{eq:UL}
    S = S_1^T \,S_2 \, S_1 \,,
\end{equation}
where the subscripts `$1$' and `$2$' refer to the mantle (one layer)
and core. $S_1$ and $S_2$ can be calculated using the formalism
described in the previous sections. Since the core density profile is
symmetric, the corresponding integral $\Delta J_2$ vanishes, whereas
$\Delta I_2$ can be calculated in full analogy with the derivation of
$\Delta I$ in Sec.~\ref{sec:mantle}. In particular, we can approximate
the core density profile by a parabola and obtain for $\Delta I_2$ an
expression which is completely analogous to that in
Eq.~\eqref{eq:phi-int}: $\Delta I_2 = \Delta I (\phi_2, V_2'' L_2 )$.
Here $\phi_2 \equiv \bar\omega_2 L_2$ is the phase acquired in the
core layer. The expression for $S_2$ is therefore
\begin{equation} \label{eq:Ucore}
    S_2 = \cos\varepsilon_2 \, \bar{S}_2
    - i \, \sin\varepsilon_2
    \begin{pmatrix}
	\sin2\bar\theta_2 & \hphantom{-}\cos2\bar\theta_2 \\
	\cos2\bar\theta_2 & -\sin2\bar\theta_2
    \end{pmatrix},
    \qquad
    \varepsilon_2 = \sin2\bar\theta_2 \, \Delta I_2 \,.
\end{equation}

For the neutrino trajectories that cross the Earth's core, it is
convenient to approximate the density profile within each mantle layer
by a linear function of the coordinate:
\begin{equation}
    V_1(z) = \bar V_1 + \Delta V_1(z),
    \qquad
    \Delta V_1(z) \approx V'_1 \frac{z}{L_1} \,,
\end{equation}
where $L_1$ is the length of one mantle layer. The advantage of this
parametrization is that $\Delta V_1(z)$ is an antisymmetric function
of $z$, and therefore $\Delta I_1$ vanishes. For $\Delta J_1$ we
obtain from Eq.~\eqref{eq:DeltaVcs}:
\begin{equation} \label{eq:DJm}
    \Delta J_1 = V'_1 L_1 \,
    \frac{\sin\phi_1 - \phi_1 \cos\phi_1}{4\phi_1^2} \,
\end{equation}
where $\phi_1 \equiv \bar\omega_1 L_1$ is the phase acquired in each
mantle layer. Then $S_1$ is given by
\begin{equation} \label{eq:Umantle}
    S_1 = \cos\varepsilon_1 \, \bar{S}_1
    + \sin\varepsilon_1
    \begin{pmatrix}
	0 & -1 \\
	1 & \hphantom{-}0
    \end{pmatrix},
    \qquad
    \varepsilon_1 = \sin2\bar\theta_1 \, \Delta J_1 \,.
\end{equation}
At first-order in $\varepsilon_1$:
\begin{equation} \label{eq:Umantle2}
    S_1 \approx
    \begin{pmatrix}
	\cos\phi_1 + i \cos2\bar\theta_1 \, \sin\phi_1
	& \sin2\bar\theta_1 \LT( -i \sin\phi_1 - \Delta J_1 \RT)
	\\
	\sin2\bar\theta_1 \LT( -i \sin\phi_1 + \Delta J_1 \RT)
	& \cos\phi_1 - i \cos2\bar\theta_1 \, \sin\phi_1
    \end{pmatrix}.
\end{equation}

In Fig.~\ref{fig:profile} we show the dependence of ($\bar{V}_1$,
$V'_1$) and ($\bar{V}_2$, $V''_2$) on the nadir angle $\Theta_\nu$.
With these functions, one can find from Eqs.~\eqref{eq:phi-int} and
Eqs.~\eqref{eq:DJm} the quantities $\Delta I_2$ and $\Delta J_1$ and
then from Eqs.~\eqref{eq:Ucore} and~\eqref{eq:Umantle} the evolution
matrices $S_1$ and $S_2$. Substituting the results into
Eq.~\eqref{eq:UL}, one obtains the evolution matrix for the whole
trajectory.

In the limit $|\Delta J_1| \ll |\sin\phi_1|$ Eq.~\eqref{eq:Umantle2}
can be rewritten as
\begin{equation}
    S_1 = D_1\,\bar{S}_1\ D_1^* \,, \qquad
    D_1 \equiv
    \begin{pmatrix}
	e^{-i\tau_1/2} & 0 \\
	0 & e^{+i\tau_1/2}
    \end{pmatrix} \,, \qquad
    \tau_1 = \arcsin \LT( \frac{\Delta J_1}{\sin\phi_1} \RT) \,,
\end{equation}
and since $\Delta J_1$ is real, $D_1$ is a pure phase matrix. In this
limit the total evolution matrix Eq.~\eqref{eq:UL} takes the form
\begin{equation} \label{eq:present}
    S = \bar{S}_1 \, D_1 \, S_2 \, D_1 \, \bar{S}_1 \,
\end{equation}
where we have omitted the two outer matrices $D_1^*$ are irrelevant
for the calculation of $P_A = |S_{12}|^2$. The two matrices $D_1$ in
Eq.~\eqref{eq:present} can be combined with $S_2$ to construct an
effective core matrix, which takes into account all the first-order
corrections (both in the core and in the mantle):
\begin{equation}
    S \equiv \bar{S}_1 \, S'_2 \, \bar{S}_1 \,, \qquad
    S'_2 \equiv D_1 \, S_2 \, D_1 \,.
\end{equation}
An advantage of this approach is that both $\bar{S}_1$ and $S'_2$ are
\emph{symmetric} matrices, so that all the results derived previously
in the constant-density approximation can be improved to take into
account the first-order corrections by simply replacing $\bar{S}_2 \to
S'_2$. The asymmetry of the mantle profile is effectively resolved.

In Fig.~\ref{fig:apx-impro} we compare the $P_A$ oscillograms obtained
by numerical calculations for the PREM matter profile with those
obtained using Eqs.~\eqref{eq:UL}, \eqref{eq:Ucore}
and~\eqref{eq:Umantle}. As can be seen from this figure the degree of
accuracy increases drastically once the corrections described here are
included.

An advantage of the described approximate analytic approach is that
the coefficients $\bar{V}$ and $V''_1$ depend solely on the nadir
angle of the neutrino trajectory $\Theta_\nu$. In particular, they do
not depend on neutrino energy and on the values of $\Dmq$ and
$\theta$.


\section{Dependence of oscillograms on 1-3 mixing, density profile and
  flavor channel}
\label{sec:channels}


\subsection{Dependence of oscillograms on 1-3 mixing}
\label{sec:13dep}

As can be seen in Fig.~\ref{fig:theta13}, with increasing $\sin^2
2\theta_{13}$ the oscillation probability increases everywhere in the
$(E_\nu ,\, \Theta_\nu)$ plane. The evolution of the oscillation
pattern appears as a ``flow'' of higher probability along nearly fixed
curves towards larger values of $\Theta_\nu$: The flow is along the
curves of the phase condition~\eqref{eq:phase1} for the mantle
trajectories, and along the curves of the collinearity condition for
the core-crossing trajectories. These lines of flow move only weakly
with $\theta_{13}$.

The region of sizable oscillation probability, $P_A \geq 1/2$, appears
first for $\sin^2 2\theta_{13} \approx 0.009$, at $\Theta_\nu =
0^\circ$ and $E_\nu = 2.8$ GeV. It is located on the parametric ridge
$A$. This large probability is due to the parametric enhancement of
the oscillations. With increase of $\sin^2 2\theta_{13}$ the region of
sizable probability expands along ridge $A$, and at $\sin^2
2\theta_{13} \approx 0.025$ reaches the inner-mantle trajectories. At
$\sin^2 2\theta_{13} \approx 0.04$ the regions of $P_A \geq 1/2$
appear also (at $\Theta_\nu = 0^\circ$) in the core MSW resonance
region and in the parametric ridge $B$. The curves of zero oscillation
probability change only very slightly with $\theta_{13}$.
The nature of the increase of the oscillation probability with
increasing $\theta_{13}$ is different in different regions of the
parameter space.

Analytic expressions for various structures in the oscillograms,
derived in Sec.~\ref{sec:oscillogr}, allow one to understand this
evolution. Two features determine dependence of the oscillograms on
1-3 mixing:
\begin{myitemize}
  \item factorization of the $\theta_{13}$ dependent factors in the
    probability, and

\item dependence of the amplitude and phase conditions on 1-3 mixing.
\end{myitemize}

Let us consider first the regions outside the resonances: $E_\nu > E_R
+ \Delta E_R^\text{max}$, and $E_\nu < E_R - \Delta E_R^\text{max}$.
Here $\Delta E_R^\text{max} = E_R \tan 2 \theta_{13}^\text{max}$ and
$\tan 2 \theta_{13}^\text{max} \sim (0.1 - 0.2)$ corresponds to the
maximal allowed values of 1-3 mixing. In practice, for the
mantle-crossing trajectories these are the regions with $E_\nu < 5$
GeV and $E_\nu > 8$ GeV. In these regions, the mixing parameter in
matter can be approximated by
\begin{equation} \label{eq:mix-app}
    \sin^2 2\theta_m \approx
    \frac{\sin^2 2\theta_{13}}{\LT|1 - \frac{2VE_\nu}{\Dmq} \RT|^2},
\end{equation}
Moreover, the half-phase
\begin{equation} \label{eq:phase-app}
    \phi \approx \frac{\Dmq}{4E_\nu}
    \LT|1 - \frac{2VE_\nu}{\Dmq} \RT| L
\end{equation}
does not depend on 1-3 mixing.
So, for the mantle domain trajectories the lines of constant phase
that determine the lines of flow do not depend on 1-3 mixing. As
follows from~\eqref{eq:mix-app} and~\eqref{eq:phase-app}, the
oscillation probability for one layer (mantle) factorizes:
\begin{equation}
    P_A \approx \sin^2 2\theta_{13} \sin^2 \phi
    \frac{1}{\LT|1 - 2 V E_\nu / \Dmq \RT|^2}.
\end{equation}
Therefore, the probability increases uniformly in the whole this area
and the lines of zero and maximum transition probability do not move.
For the core crossing trajectories similar analysis holds for $E_\nu <
2$ GeV.

Let us first consider the parametric resonance
condition~\eqref{eq:parcond} that determines approximately the lines
of ``flow''. Beyond the MSW resonance regions, the phases $\phi_i$,
and therefore $c_i$ and $s_i$ ($i = 1,2$), do not depend on the 1-3
mixing. The mixing angles in matter enter the condition as cosines
$\cos 2\theta_i$. Outside the resonance regions $\theta_i \approx
\theta_{13} \ll 1$ or $\theta_i \approx \pi/2$ and therefore $\cos
2\theta_i \approx \pm 1$ weakly depends on 1-3 mixing. So, the
condition (and therefore the lines of flow) shifts only weakly with
change of $\theta_{13}$.

Let us now consider the resonance regions $(2 < E_\nu < 12)$ GeV. Here
dependence of the transition probability on $\sin^2 2\theta_{13}$ is
non-linear. Not only the depth of the oscillations, but also the
oscillation length in matter in each layer depends on $\sin^2
2\theta_{13}$ substantially, and the latter influences the
interference effects. As a result, the transition probability changes
with $\theta_{13}$ differently in different regions, and in addition
the ``lines of flow'' shift.
\begin{myitemize}
  \item The MSW resonance peak in the mantle is determined by the
    condition~\eqref{eq:Eres}.  With increasing $\sin^2 2\theta_{13}$,
    the peak shifts to larger values of $\Theta_\nu$ and to slightly
    larger energies. This can be readily understood. Indeed, the
    oscillation length at the resonance decreases with increasing
    $\theta_{13}$ as $l_m = l_{\nu}/\sin 2\theta_{13}$ and therefore
    the condition $\phi_1 = \pi/2$ is satisfied for shorter (more
    external) trajectories. For these trajectories the average density
    becomes smaller, so that the resonance energy increases: $E_R
    \propto 1/\bar{V}$. The width of the resonance peak at half height
    increases as $\tan 2 \theta_{13}$ both in neutrino energy and in
    $\Theta_\nu$ variables (see the discussion below
    Eq.~\eqref{eq:Eres}).
    Using the resonance condition~\eqref{eq:Eres}, we obtain
    from~\eqref{eq:phase2}
    \begin{equation} \label{eq:theta13det}
	\tan 2\theta_{13} = \frac{\pi}{ 2 \bar{V}(\Theta_\nu) R
	  \cos\Theta_\nu} \equiv\frac{\pi}{d}\,,
    \end{equation}
    where $d$ is the column density that corresponds to maximum of the
    transition probability. Eq.~\eqref{eq:theta13det} gives an
    immediate relation between $\theta_{13}$ and the nadir angle of
    the neutrino trajectory on which the absolute maximum of the
    transition probability is realized. This can, in principle, be
    used for measuring $\theta_{13}$: the method would simply consist
    in the determination of $\Theta_\nu$ of the mantle-only crossing
    trajectory corresponding to the absolute maximum of the conversion
    probability $P_A=1-P_{ee}$.
    
  \item Core ridge slightly shifts with increase of $\theta_{13}$ to
    larger energies, especially at $\Theta_\nu \approx 0$.
    
  \item For ridge $A$, the region of sizable transition probability
    and the position of the maximum also move towards smaller $|\cos
    \Theta_\nu|$. With increasing $\theta_{13}$ the oscillation length
    decreases, especially in the resonance region. Therefore, the same
    phases in the core and mantle can be obtained for shorter
    trajectories in the core and therefore for larger $\Theta_\nu$.
    
  \item Ridge $B$ evolves weaker with $\theta_{13}$: the energy
    corresponding to the maximum of the transition probability stays
    rather close to that of the mantle MSW resonance, $E_\nu \approx
    (5 - 6)$ GeV. At higher energies the ridge is in the factorization
    region. With increase of $\theta_{13}$ at $\Theta_\nu \approx 0$
    the ridge shifts to smaller $E_\nu$.
   
   \item The situation for ridge $C$ is similar to ridge $B$.
\end{myitemize}


\subsection{Dependence on the Earth's density profile}
\label{sec:prof}

In different parts of the $(E_\nu ,\, \Theta_\nu)$ plane the
oscillation probabilities have different sensitivity to the
modifications of matter density profile. The sensitivity is very weak
(independently of the form of perturbation) in the following parts:
\begin{myitemize}
  \item $\Theta_\nu > 84^\circ$: here  the length of the trajectory,
    and therefore the oscillation phase, are small, so that effect of
    ``vacuum mimicking''~\cite{Wolfenstein:1977ue, vacmim} takes
    place. To a good approximation matter does not affect the
    oscillation probabilities, irrespectively of whether or not the
    matter-induced potential $V$ is small compared to the kinetic
    energy difference $\Dmq_{31}/2E_\nu$.
    
  \item  $E_\nu < 2$ GeV: here  one has $V\ll \Dmq_{31}/2E_\nu$, and so
    matter effects on the oscillations driven by 1-3 mixing and
    splitting are small for all values of $\Theta_\nu$.
\end{myitemize}

The region of high sensitivity to the density profile of the Earth is
bounded by $E_\nu > (3 - 4)$ GeV and $\Theta_\nu < 66^\circ$. This is
the region of energies close to and above the resonance energies
(matter dominance) and of sufficiently long trajectories. The latter
condition ensures an accumulation of the matter effects over long
distances and therefore a sensitivity to large scale structures of the
density profile. For the core-crossing trajectories the border of the
sensitivity region is lower: $E_\nu \simeq 2$ GeV. Changes of the
oscillograms in this region depend on the specific form of the
modification of the matter density distribution.

For illustration, we consider here the effects of three different
modifications of the Earth density profile: (1) replacing PREM profile
by constant-density mantle and core layers, (2) modification of the
core/mantle density ratio, and (3) changes of the position of the
border between the mantle and the core.

\textitem{1.} In Fig.~\ref{fig:apx-fixed} we show the results of the
fixed constant-density layers approximation, characterized by the
constant potentials $V_1$ and $V_2$ in the mantle and core. This
profile can be considered as an extreme case of flattening of the
mantle and core density distributions. For the mantle-only crossing
trajectories there are two lines in the oscillograms where the
oscillation probabilities for the two profiles are equal: (i)
$\Theta_\nu \approx 53^\circ$, which corresponds to the trajectory
with $\bar{V}_{\rm PREM} = {V}_1$, and (ii) the resonance energy curve
$E_\nu \simeq E_R(\Theta_\nu)$. Indeed, along the resonance curve the
first-order correction to $P_A$ due to the deviation of the density
profile from the averaged constant one disappears (see the discussion
in Sec.~\ref{sec:mantle}).

For $\Theta_\nu < 53^\circ$ the oscillation pattern for the fixed
constant-density profile is shifted to higher energies compared to
that for the PREM profile. Furthermore, for $E_\nu > E_R(\Theta_\nu)$
one has $P_\text{const} > P_\text{PREM}$, whereas for lower energies,
$E_\nu < E_R(\Theta_\nu)$, $P_\text{const} < P_\text{PREM}$, which
essentially reflects the shift of the resonance peak when profile is
changed. The shift of contours in the energy scale increases as
$\Theta_\nu$ decreases and reaches maximum for the deepest mantle
trajectories.  Quantitatively, in the high energy region, $E_\nu > 6$
GeV, for $\Theta_\nu \sim 37^\circ$ the changes of probability equal
$P_\text{const}/P_\text{PREM} = 0.2/0.06$, $0.3/0.1$, $0.4/0.2$,
$0.5/0.3$, $0.9/0.5$. That is, the modification is characterized by
factor 2 - 3. The shift of contours in energy scale is about 1 GeV at
$E_\nu \sim 10$ GeV. The differences between the results for the two
profiles weakly depend on $\theta_{13}$. For $\Theta_\nu > 53^\circ$,
the PREM profile gives higher probabilities than those for the fixed
constant-density layers approximation for $E_\nu > E_R(\Theta_\nu)$
and lower probability for $E_\nu < E_R(\Theta_\nu)$.

For core crossing trajectories the size of changes is similar.

\textitem{2.} Fig.~\ref{fig:scale} illustrates the effects of the
increase (decrease) of the core density: $V_2(x) \rightarrow k
V_2(x)$, where $k = const$. The shapes of the density profiles in the
mantle and core are taken according to the PREM profile. The total
mass of the Earth is unchanged and therefore the density of the mantle
should be reduced (increased) correspondingly: $\Delta V_1/ V_1 = -
[(R/R_c)^3 - 1]^{-1} (\bar{V}_2 /\bar{V}_1)\Delta V_2 /V_2 \approx -
0.3 \Delta V_2 / V_2$.
The effects of the increase and decrease of the core potential on the
oscillogram are opposite and for definiteness we will consider the
case of an increase of $V_2$, so that for mantle-only crossing
trajectories $V_1$ decreases. Since the resonance energy is
proportional to $1/V$, a decrease of density leads to a shift of the
resonance peak to higher energies and also to smaller $\Theta_\nu$, to
satisfy the phase condition $\phi = \pi/2$. The strongest effect is
for $\Theta_\nu < 60^\circ$ and above the resonance. Quantitatively, a
$\sim 6\%$ decrease of the mantle density (corresponding to the $20\%$
increase of $V_2$) leads, for $\Theta_\nu < 37^\circ$, to an about 1
GeV upward shift of the contour curves. As follows from the figure,
the increase of probabilities with respect to those for the PREM
profile is $P_\text{20\%}/P_\text{PREM}=0.1/0.03$, $0.2/0.1, 0.4/0.2$,
$0.5/0.3, 0.7/0.5$. The curve of unchanged probability is close to the
resonance curve. Below the resonance energy the PREM profile leads to
larger probabilities than the modified one.

Notice that these changes are rather similar to the changes for the
constant-density layers profile. The reason is that in both cases in
the inner parts of the mantle the density is smaller than that given
by the PREM density profile.

The effect of density modifications on the transition probability is
nearly linear in the variable of density change.

For core-crossing trajectories, an increase of the core density leads
to a shift of the oscillatory pattern to larger $\Theta_\nu$ and
higher energies, and also results in an increase of probability. These
features can be understood using the parametric resonance condition.

\textitem{3.} Fig.~\ref{fig:shift} illustrates the effects of increase
(decrease) of the core radius. Again, the requirement of keeping the
total mass of the Earth fixed imposes a rescaling of the overall core
and mantle densities. As in the previous case, this rescaling is what
induces the most important effects on the oscillogram.

\FIGURE[!t]{
  \includegraphics[width=145mm]{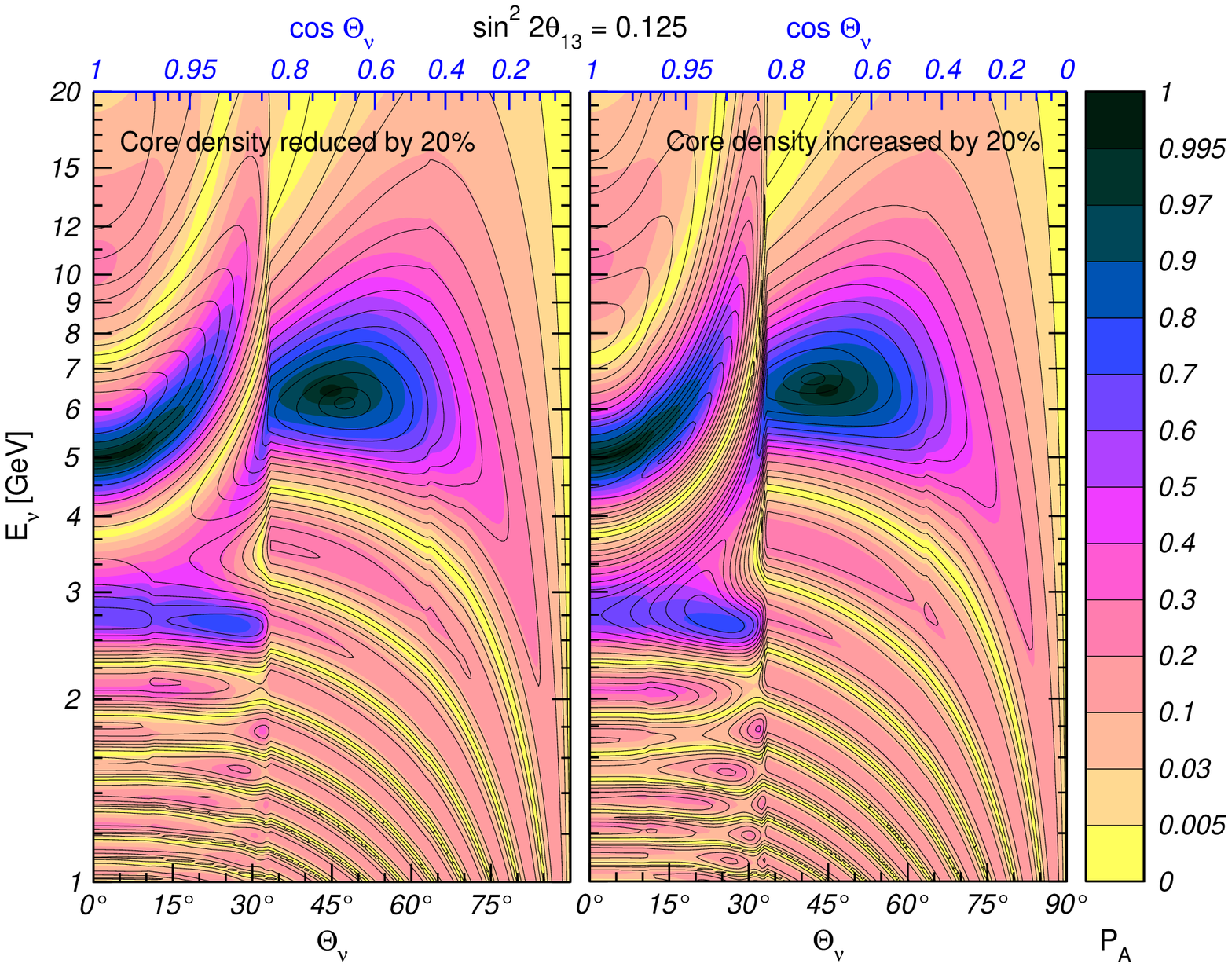}
  \caption{\label{fig:scale}%
    $P_A$ oscillograms for the PREM density profile (colored regions)
    and for a 20\% variations of the core/mantle density ratio. The
    total mass of the Earth is kept fixed.}
  }

\FIGURE[!t]{
  \includegraphics[width=145mm]{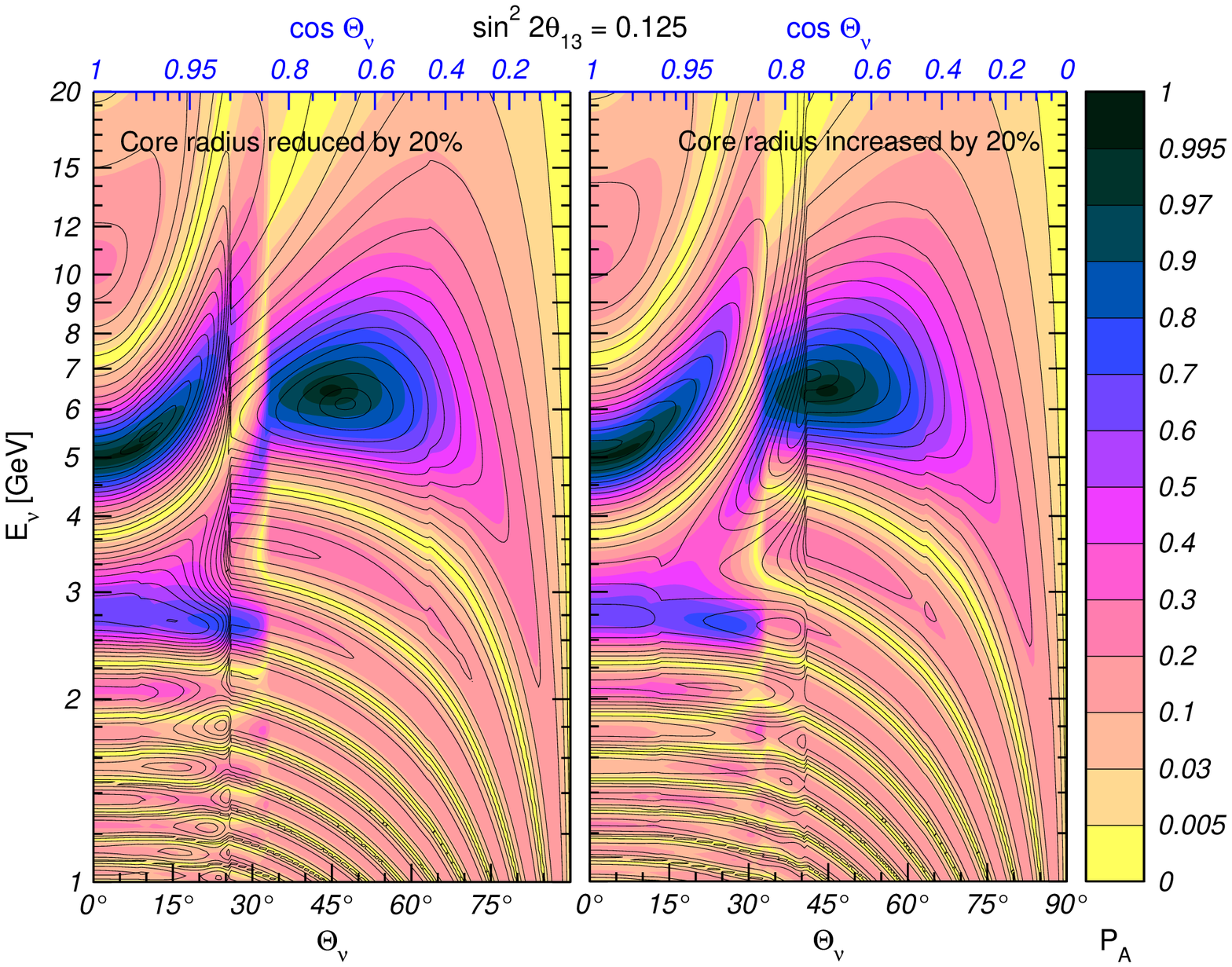}
  \caption{\label{fig:shift}%
    $P_A$ oscillograms for the PREM density profile (colored regions)
    and for a 20\% variations of the core radius. The total mass of
    the Earth is kept fixed.}
  }

The sensitivity of the oscillation probabilities to the variations of
the matter density distribution in the Earth can in principle be used
for studying the Earth interior with neutrinos, \ie\ to perform an
oscillation tomography of the Earth~\cite{AMS2}.
Using the features described above one can work out the criteria for
the selection of events that are sensitive to a given type of
variations of the profile. For instance, in the case of flattening of
the density distribution, one can divide the whole parameter space
into 4 parts:
(1) $\Theta_\nu < 53^\circ$, $E_\nu > E_R$;
(2) $\Theta_\nu > 53^\circ$, $E_\nu < E_R$;
(3) $\Theta_\nu < 53^\circ$, $E_\nu < E_R$;
(4) $\Theta_\nu > 53^\circ$, $E_\nu > E_R$,
and select a sample of the $e$-like events sensitive to these regions.
The flat profile gives larger number of events then PREM profile in
(1) and (2) regions and smaller number of events in the regions (3)
and (4). This feature allows one to distinguish effect of flattening
from other possible modifications of the profile as well as effect of
uncertainty in $\theta_{13}$.

Notice that the changes of the oscillation probability due to
modifications of the core/mantle density ratio and of the position of
the border between the mantle and the core can be noticeable if the
magnitude of the modifications is as large as $20\%$ (see
Figs.~\ref{fig:scale} and~\ref{fig:shift}). However, to detect them
experimentally, the energy and nadir angle resolutions should be high
as well as large event statistics is necessary. Therefore, the
oscillation tomography of the Earth will put very challenging demands
to future experiments.


\subsection{Probabilities for other oscillation channels}

In Figs.~\ref{fig:osc-050} and~\ref{fig:osc-125} we show the
oscillograms for the other channels. As follows from the discussion in
Sec.~\ref{sec:heosc}, in the approximation of zero 1-2 splitting
($\Dmq_{21} = 0$) all the probabilities that involve $\nu_e$ depend on
the single probability $P_A$. This is related to the fact that $\nu_e$
is unchanged upon going from the flavor basis to the propagation one.

The probability $P_{ee}=1-P_A$ is just complementary to $P_A$ with all
the features inverted. $P_{e\mu}$ is just $P_A$ scaled by the factor
$s^2_{23}$ (see the middle panel in Fig.~\ref{fig:osc-050}). The
maximal value of this transition probability is therefore $s^2_{23}$.
Similarly, $P_{e\tau}$ is $P_A$ scaled by the factor $c^2_{23}$.

\PAGEFIGURE{
  \includegraphics[width=145mm]{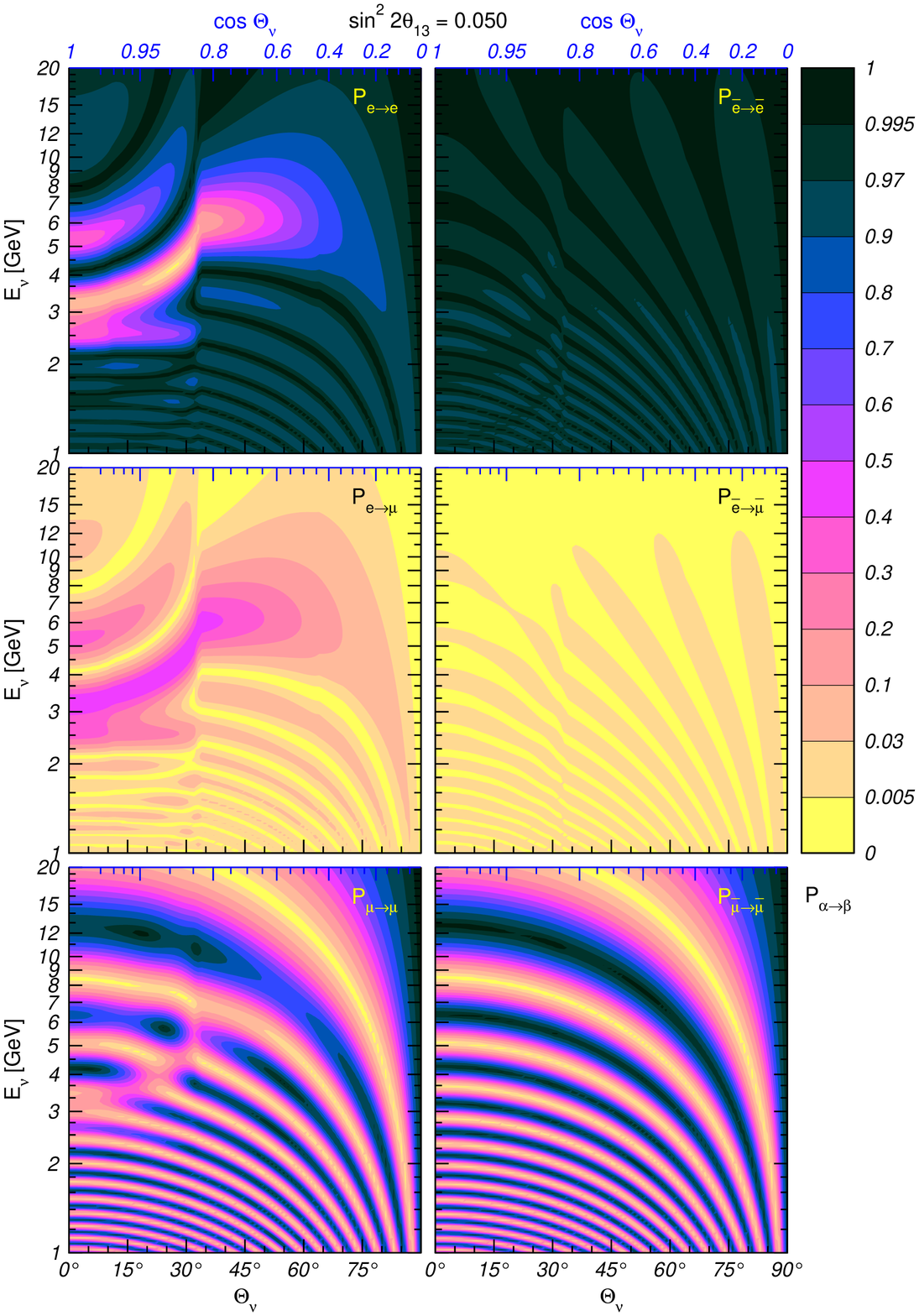}
  \caption{\label{fig:osc-050}%
    $P_{ee}$, $P_{e\mu}$ and $P_{\mu\mu}$ oscillograms, for neutrinos
    (left panels) and antineutrinos (right panels),
    $\sin^22\theta_{13} = 0.05$, $\sin^22\theta_{23} = 1$ and
    $\Dmq_{21} = 0$.}
  }

\PAGEFIGURE{
  \includegraphics[width=145mm]{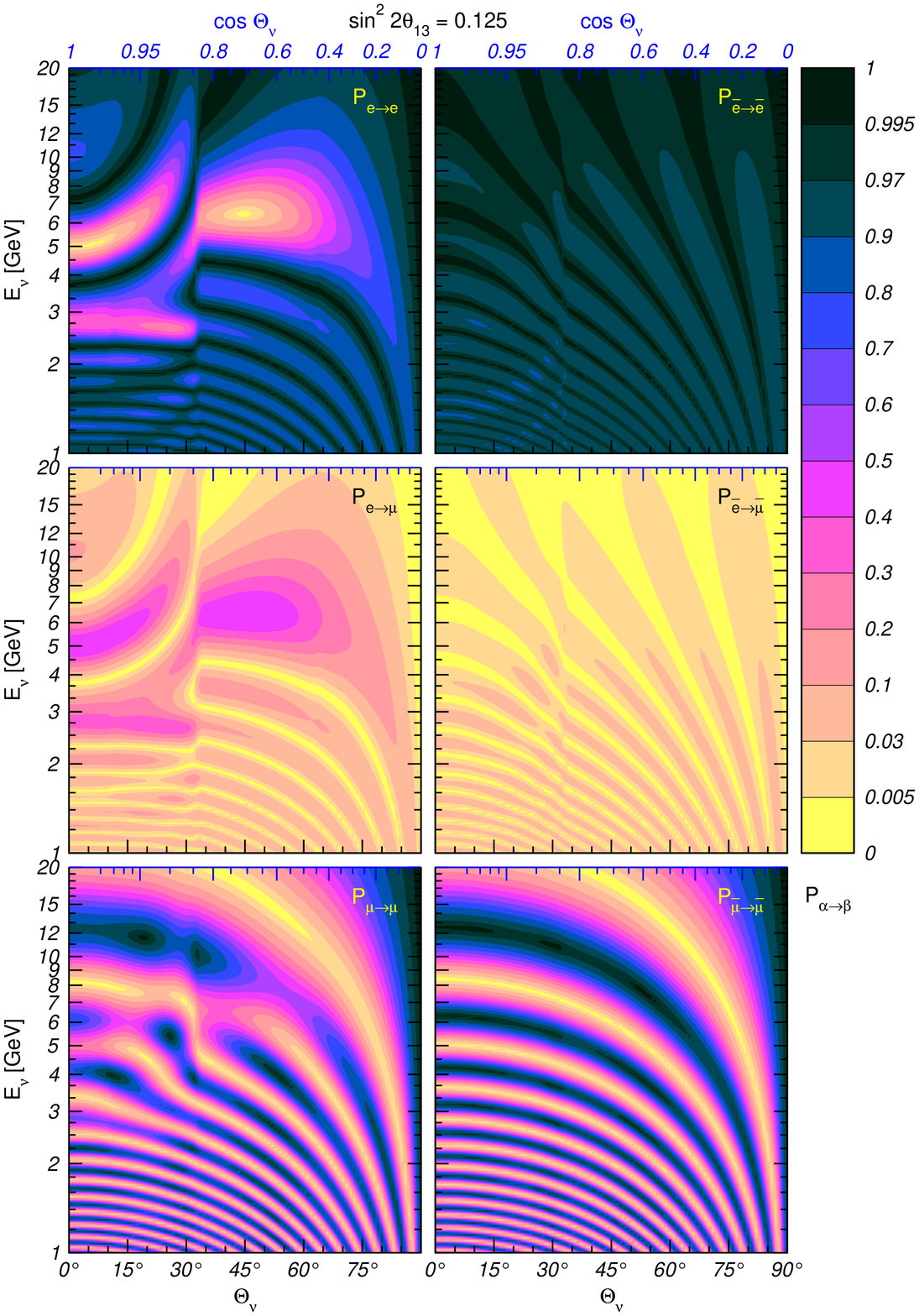}
  \caption{\label{fig:osc-125}%
    The same as in Fig.~\ref{fig:osc-050} but for $\sin^22\theta_{13}
    = 0.125$.}
  }

The probabilities of transitions that do not involve $\nu_e$ have more
complicated structure since now both the initial and the final states
do not belong to the propagation basis and therefore some additional
interference occurs. The survival probability $P_{\mu\mu}$ in
Eq.~\eqref{eq:prob-mumu} can be written as
\begin{equation} \label{eq:mumupr}
    P_{\mu\mu} = P_{vac} - s^4_{23} P_A
    + 2 s^2_{23} c^2_{23} \, (\Re A_{33} - \cos 2\phi_{vac})\,,
\end{equation}
where $P_{vac} = 1 - \sin^2 2\theta_{23}\sin^2 \phi_{vac}$ is the
usual 2-flavor vacuum oscillation probability and $\phi_{vac} \equiv
\Dmq_{31} L/4E_\nu$ is the vacuum oscillation phase. $P_{vac}$
describes the oscillation effect in the absence on the 1-3 mixing. The
other two terms in~\eqref{eq:mumupr} describe the effects of the 1-3
mixing. We can use the approximate results of Sec.~\ref{sec:approx} to
estimate $\Re A_{33}$. Recall that in Sec.~\ref{sec:approx} we used a
symmetric Hamiltonian for $2\nu$ system, which differs from the one
introduced in Sec.~\ref{sec:3fosc} by a term proportional to the unit
matrix~\eqref{eq:HHconnect}. The term $\Re A_{33}$ essentially
reflects evolution of the 1-3 system with respect to the state
$\tilde{\nu}_2$ and so the $3\nu$ form of the Hamiltonian should be
restored. According to Eq.~\eqref{eq:HHconnect} the relation between
the corresponding evolution matrices (up to the removed decoupled
state) is
\begin{equation}
    \label{eq:s-relation}
    \tilde{S} = e^{-i\psi} S, \qquad \psi \equiv
    \int_0^L \LT(\frac{\Dmq_{31}}{4E_\nu} + \frac{V}{2} \RT) dx \,.
\end{equation}
Then
\begin{equation} \label{eq:a33}
    A_{33} = e^{-i\psi}\,S_{22}\,,
\end{equation}
where $S_{22}$ is the 22 element of the matrix~\eqref{eq:Ubar} with
the correction given in~\eqref{eq:DeltaU}.

Let us consider the mantle trajectories. Then from~\eqref{eq:a33} we
get explicitly
\begin{equation}
    \Re A_{33} =
    \cos (\phi + \psi) + 2 \sin^2 \bar{\theta}_{13} \sin\psi [\sin\phi
    + 2 \cos^2 \bar{\theta}_{13}  \Delta I ] \,,
\end{equation}
where the phase $\phi$ and the correction integral $\Delta I$ are
defined in~\eqref{eq:Ubar} and~\eqref{eq:DeltaVcs}, respectively, and
the angle $\bar{\theta}_{13}$ is the 1-3 mixing angle in matter
calculated at the average value of the matter density. Then the
survival probability $P_{\mu\mu}$ for the mantle-only crossing
trajectory can be written as
\begin{multline} \label{eq:mumucor}
    P_{\mu\mu} =
    1 - \sin^2 2\theta_{23}\sin^2 \LT( \frac{\phi + \psi}{2}\RT)
    - s_{23}^4 P_A
    \\
    + \frac{1}{2} \sin^2 2\theta_{23} \sin \psi
    \LT( 2 \sin^2 \bar{\theta}_{13} \sin\phi
    + \sin^2 2\bar{\theta}_{13}\, \Delta I\RT) \,.
\end{multline}

Notice that the first two terms here correspond to 2-flavor vacuum
probability with the modified phase. The probability can be rewritten
as
\begin{equation}
    P_{\mu\mu} = P_{vac} + \Delta P
\end{equation}
with
\begin{multline} \label{eq:mumucor2}
    \Delta P = - s^4_{23} P_A - 2 s^2_{23} c^2_{23}
    \big[\cos 2\phi_{vac} - \cos (\phi + \psi)
    \\[1mm]
    - 2\sin^2 \bar{\theta}_{13} \sin \psi\sin\phi
    - \sin^2 2\bar{\theta}_{13}\sin \psi\, \Delta I \big].
\end{multline}
In the limit $s_{13} \to 0$ one has
\begin{equation}
    \phi \approx \int_0^L \LT(\frac{\Dmq}{4E_\nu} -
    \frac{V(x)}{2}\RT) dx
\end{equation}
which, together with~\eqref{eq:s-relation}, implies $\phi + \psi =
2\phi_{vac}$.

These formulas can be used for the analysis of numerical results shown
in Figs.~\ref{fig:osc-050} and~\ref{fig:osc-125}. The first correction
term in Eq.~\eqref{eq:mumucor2} is negative, so that it reduces the
survival probability. The strongest effect of this mixing occurs in
the resonance region. For high energies the 1-3 mixing is suppressed
and $P_{\mu\mu}$ is again described well by the vacuum oscillation
formula.
Notice that in some regions with zero $P_{vac}$ the 1-3 mixing leads
to the positive contribution which is due to the last term
in~\eqref{eq:mumupr} so that the lines of of zero probability are
interrupted. Also there are no lines of maximal probability, as zero
lines for $P_A$.

All the correction terms to $P_{vac}$ are in general of the same
order; therefore the modification of the vacuum probability is rather
complex and not immediately related to $P_{e\mu}$. The corrections are
in general large in the regions of the parameter space where
$P_{e\mu}$ is large, \ie\ in the resonance regions and in the places
of the resonance peaks and parametric ridges of $P_A$. Instead of
continuous lines of maximal probability, in the case of non-zero 1-3
mixing one obtains saddle points and local maxima. In particular, the
saddle point of the $P_{\mu\mu}$ probability appears in the mantle
resonance region.

The structure of the transition probability $P_{\mu\tau}$ is similar
to that of $P_{\mu\mu}$.

In the antineutrino channel, the 2-flavor probability $\bar{P}_2$ is
suppressed by matter and the strongest transitions occur in the vacuum
oscillation region. Some interference effects are seen for the core
domain but no substantial parametric enhancement is realized.


\section{Discussion and conclusions}
\label{sec:conclusions}

\noindent\textbf{1.} We have worked out a detailed and comprehensive
description of neutrino oscillations driven by non-zero 1-3 mixing 
inside the Earth. The description is given in terms of the
oscillograms of the Earth: contours of constant oscillation
probabilities in the $(E_\nu ,\, \Theta_\nu)$ plane. In this first
publication we have neglected the 1-2 mass splitting $\Dmq_{21}$,
which is a good approximation for high neutrino energies, $E_\nu >
1-2$ GeV.

\textitem{2.} We found that the oscillograms have a regular structure
with several generic features: (i) the MSW peak in the mantle domain
of the oscillogram; (ii) the MSW peak (ridge) in the core domain,
(iii) three parametric ridges in the core domain; (iv) regular
oscillatory pattern at low energies that has different features in the
core and in the mantle domains of the oscillograms. We presented a 
detailed description of these structures: their position in the 
$(E_\nu ,\, \Theta_\nu)$ plane, their evolution with changing 1-3 
mixing and their dependence on the density profile of the Earth.

The most interesting features of the oscillograms appear at relatively
high energies, $E_\nu = (2 - 12)$ GeV, and $\Theta_\nu < 75^{\circ}$,
\ie\ in the resonance region. Notice that this region is not covered
by the existing or forthcoming accelerator experiments, and on the 
other hand, statistics in the current atmospheric neutrino experiments
is too low. All these experiments can only study very small effects on
the ``tails'' of those interesting oscillation phenomena. Exploration 
of the resonance regions thus constitutes a significant experimental 
challenge for future experiments.

\textitem{3.} We studied the accuracy of the calculations based on the
constant-density layers approximation to the Earth matter density
profile. This approximation reproduces all the features of the 
oscillograms qualitatively well, though there are some quantitative
differences and shifts of the structures in the $(E_\nu ,\,
\Theta_\nu)$ plane. The strongest deviations appear in the domain of
deep mantle trajectories and high energies ($E_\nu > 6$ GeV).

\textitem{4.} We showed that a complete physics interpretation of the 
oscillograms can be given in terms of different realizations of just
two conditions: (i) the amplitude (or resonance) condition and (ii)
the phase condition. In the case of one layer of constant density
these conditions are reduced to the MSW resonance condition and
half-phase equality $\phi = \pi/2 + \pi k$. They determine the
position of the absolute maxima of the transition probability. For 
three layers of constant densities the amplitude condition is reduced 
to the parametric resonance condition. This condition describes the
position of the parametric ridges. The phase condition gives the
position of the maximum along the ridge.

We show that the parametric resonance condition formulated for two
layers describes the extrema and the ridges in the three layers case. 
This is a consequence of the symmetry of the overall matter density 
profile.
In the case of three layers, the amplitude and the phase conditions
determine not only the absolute maxima of the transition probability 
(as in the one-layer case), but also its local maxima and saddle 
points.

\textitem{5.} We generalized the amplitude and phase conditions to the
case of varying densities in each layer. The generalization is not
unique. In this connection we introduced the generalized resonance 
condition and the collinearity condition. The two conditions coincide 
in the case of constant density layers with symmetric overall profiles
but differ in the non-constant density case. We showed that both these
generalized conditions describe the positions of various structures of
the oscillograms, in particular, of the extrema, very accurately.

\textitem{6.} We derived approximate analytic formulas for the
probabilities. For this we have developed a perturbation theory in
deviations from the constant density $\Delta V / \bar{V}$. We showed
that already the first order approximation in $\Delta V / \bar{V}$
reproduces the oscillograms for realistic (PREM) density profile of
the Earth with a high precision. Again, the symmetry of the density
profile plays the key role.

\textitem{7.} We studied the dependence of the oscillograms on
$\theta_{13}$. The changes of the oscillograms with increasing 
$\theta_{13}$ have a character of flow of high probabilities towards
the regions of larger $\Theta_\nu$. The lines of flow shift only 
weakly with changing $\theta_{13}$. We found that the transition 
probability $P_A$ can be of the order 1 for $\sin^2 2\theta_{13}$ as 
small as 0.01. Therefore, even if the next generation of reactor and
accelerator experiments fail to find non-zero 1-3 mixing, significant
oscillation effects due to this mixing may still show up in
atmospheric neutrino data. Those are expected in the region $E_\nu
\sim 3 - 5$ GeV and $\Theta_\nu \approx 0^\circ - 26^\circ$.

\textitem{8.} We studied the dependence of the oscillograms on the
density profile of the Earth. We found the regions in the $(E_\nu ,\,
\Theta_\nu)$ plane where the sensitivity to various perturbations of
the density profile is maximal, and we identified the corresponding
effects. In particular, the dependence of the oscillograms on
flattening of the density distributions inside the layers, on the
changes of the overall densities of the core and mantle and on the
position of the border between the mantle and the core has been
quantified. This analysis can be used for discussions of the
oscillation tomography of the Earth.

\textitem{9.} The oscillograms for different flavor channels as well 
as for neutrinos and antineutrinos have been constructed and their
properties discussed. 

\textitem{}Various applications of the results obtained in this paper
will be presented in forthcoming publications.


\acknowledgments

We would like to thank T.\ Schwetz for useful discussions and E.\ Lisi
and W.\ Winter for useful communications. EA was supported by the
Wenner-Gren Foundation as an Axel Wenner-Gren visiting professor at
the Royal Institute of Technology.


\end{document}